\documentclass[draftcls,onecolumn]{IEEEtran}

\usepackage{tikz}
\usetikzlibrary{shapes}
\usepackage{hyperref}
\usepackage{subfigure}
\usepackage{graphicx,footnote}
\usepackage{color}
\usepackage[cmex10]{amsmath}
\usepackage{amssymb,amsfonts}

\newtheorem{theorem}{Theorem}
\newtheorem{lemma}{Lemma}

\newtheorem{corollary}{Corollary}
\newtheorem{remark}{Remark}
\newtheorem{claim}{Claim}
\newtheorem{definition}{Definition}

\begin{document}

\title{Analysis of a Collaborative Filter Based on Popularity Amongst Neighbors}
\author{Kishor Barman \footnote{The work of Kishor Barman was supported by the Infosys fellowship.}\\
School of Technology and Computer Science\\
Tata Institute of Fundamental Research\\
Mumbai, India\\
Email: kishor@tcs.tifr.res.in\\
\vspace{0.2in}
Onkar Dabeer\\
School of Technology and Computer Science\\
Tata Institute of Fundamental Research\\
Mumbai, India\\
Email: onkar@tcs.tifr.res.in
}

\maketitle

\begin{abstract}
In this paper, we analyze a collaborative filter that answers the simple question: What is popular amongst your ``friends''?  While this basic principle seems to be prevalent in many practical implementations, there does not appear to be much theoretical analysis of its performance. In this paper, we partly fill this gap. While recent works on this topic, such as the low-rank matrix completion literature, consider the probability of error in recovering the {\it entire} rating matrix, we consider probability of an error in an individual recommendation (bit error rate (BER)). For a mathematical model introduced in \cite{Aditya1, Aditya2}, we identify three regimes of operation for our algorithm (named Popularity Amongst Friends ($\mathtt{PAF}$)) in the limit as the matrix size grows to infinity. In a regime characterized by large number of samples and small degrees of freedom (defined precisely for the model in the paper), the asymptotic BER is zero; in a regime characterized by large number of samples and large degrees of freedom, the asymptotic BER is bounded away from 0 and 1/2 (and is identified exactly except for a special case); and in a regime characterized by a small number of samples, the algorithm fails. We then compare these results with the performance of the optimal recommender. We also present numerical results for the MovieLens and Netflix datasets. We discuss the empirical performance in light of our theoretical results and compare with an approach (\cite{Montanari1}) based on low-rank matrix completion. 

\end{abstract}



\section{Introduction} 
Recommendation systems  suggest relevant content to users based on their previous choices. For example, it is common to predict user-item ratings based on available ratings and recommend items based on the predicted values (see \cite{netflixprize}).  In the collaborative filtering (CF) approach to recommender systems \cite{collaborative1}, information about a group of users is used to make recommendations to an individual user. There are two popular classes of CF techniques: a) neighborhood based methods (\cite{Koren3}, \cite{neighborhood1}, \cite{neighborhood_1}, \cite{186}), and b) latent factor models \cite{Koren2}. Neighborhood based methods compute similarities amongst the users (and/or amongst the items), and use information about a set of  ``similar'' users (and/or ``similar" items)  to make recommendations. On the other hand, the latent factor models assume that the entire user-item rating matrix is described by a small number of parameters, which are then estimated from available data. For example, the low-rank matrix model in \cite{Montanari1}, \cite{Candes3} is an example of this class. In the remainder of this section, we outline our goals in the context of existing works, and briefly describe the nature of our results.


\subsection{Prior Work and Our Goals}
Recently there has been a lot of interest in obtaining fundamental limits on the number of samples needed to recover a low-rank matrix with high probability  (\cite{Candes1,Montanari1,Bresler1}, \cite{Candes3}). Most of these methods try to find a matrix with lowest possible rank that agree with the observed samples. This is reminiscent of {\it compressed sensing}, where one tries to find the sparsest vector that satisfies certain affine constraints \cite{Donoho06}, \cite{Candes2}. In another model (\cite{Aditya1, Aditya2}), the rating matrix is assumed to be obtained from a block constant matrix by applying unknown row and column permutations, a noisy discrete memoryless channel representing noisy user behavior, and an erasure channel denoting missing entries. Instead of matrix completion, the goal for such a model is to  estimate the underlying ``noiseless'' matrix and the performance is dictated by the cluster size (the size of the block of constancy). For their respective models, the above listed works derive a threshold result: If the number of degrees of freedom (defined appropriately for the model) is larger than a threshold, then error free recovery is not possible, but otherwise, there is a polynomial time algorithm that recovers the entire matrix with high probability. Since empirical results (\cite{netflix-leaderboard}, \cite{neighborhood1}, \cite{dhilon1}) suggest that perfect recovery of the entire matrix might not be possible in practice, it is natural to seek a finer analysis in the regime where perfect recovery is not possible. 
In practice, we need not predict all the missing ratings - it suffices to recommend a few items with high ratings. With this in mind, in this paper we recommend one item to each user, and consider the probability that a given recommendation is incorrect as the performance metric.
 Using this metric, we seek to
develop a theoretical understanding of a basic principle that is prevalent in practical systems. This basic principle recommends items to an individual based on their popularity amongst similar users and is the main motivation for neighborhood based methods (\cite{neighborhood_1}, \cite{neighborhood1}, \cite{Koren3}). This principle is also similar to the $k$-nearest neighbors (KNN) algorithms for classification \cite{bishop1}. In this paper, we analyze a collaborative filter based on this principle for the data model proposed in \cite{Aditya1, Aditya2}. Further, we also evaluate and discuss performance on the MovieLens and Netflix datasets in light of our theoretical results and earlier works inspired by low-rank matrix completion/approximation. Below, we summarize  our main results.

\subsection{Organization and Summary of Results}

Typical rating data belongs to a finite alphabet. In this paper, we consider a binary alphabet (`like' or `dislike'), which is of special interest (see Section \ref{subsec:model} for a discussion of this point).
In Section \ref{subsec:locPop}, we describe our algorithm - named Popularity Amongst Friends ($\mathtt{PAF}$) - for a binary rating matrix. In Section \ref{subsec:exp}, we show some experimental results on the MovieLens and Netflix datasets. We compare with OptSpace \cite{Montanari1}, which is motivated by the low-rank completion problem and is a representative of this class of works. The empirical results reveal that the $\mathtt{PAF}$ algorithm has similar BER compared to OptSpace. We also present results for different values of the algorithm parameter (size of list of friends). Having demonstrated the algorithm performance on real data, in Section~\ref{sec:analysis}, we turn to its theoretical analysis. We consider the data model proposed in \cite{Aditya1,Aditya2}.

{\bf Summary of the data model:}
To motivate this model, consider an ideal situation where users and items are clustered, and users within a cluster rate items within a cluster by the same value. The rating matrix in this ideal situation (denoted by $\mathbf X$) is then a block constant matrix. The observations are obtained from $\mathbf X$ by passing its entries through a binary symmetric channel (BSC) with parameter $p$ (defined in Section \ref{subsec:model}), and an erasure channel with erasure probability $\epsilon$ (defined in Section \ref{subsec:model}). Moreover, the row and column clusters are {\em unknown}. The block constant model captures the fact that similar users rate similar items similarly, and the unknown row (column) clusters represent the fact that the sets of similar users (items) are not known. The erasures represent missing data, while the BSC represents the noisy behavior of the users. This model is described in detail in Section \ref{subsec:model}. In this paper, we present a detailed analysis of \texttt{PAF} for this model for the underlying data.

  \begin{figure}
    \begin{center}
      \begin{tikzpicture}[scale=1.5]
      \path (0,0) coordinate (origin);    
      \path (8.4,0) coordinate (x1);
      \path (0,4.4) coordinate (y1);
      
      \draw (origin) edge[->] (y1) (origin) edge[->] (x1);
      \draw (4,-0.05) -- (4,0.05);
      \draw (6,-0.3) node {$\alpha \rightarrow $};
      \draw (-0.3, 2) node[rotate=90] {$\log[\text{cluster size}] \rightarrow$};
      \draw (4,-0.3) node {1/2};
      \draw (0,-0.3) node {0};
      
      \draw (4,0) -- (4,4);

      \fill[gray!50] (origin) -- (4,4) -- (0,4) -- cycle;
      \draw (1.8,3.2) node[black, scale=1, rectangle split, rectangle split parts=2, draw,  fill=white]{
        Phase I
        \nodepart{second}
        BER=0
      };

      \fill[red!50] (origin) -- (4,0) -- (4,4) -- cycle;
       \draw (2.4,0.7) node[black, scale=1, rectangle split, rectangle split parts=2, draw,  fill=white]{
        Phase II
        \nodepart{second}
        $0<$BER$<1/2$
      };

      \fill[green!40] (4,0) -- (8,0) -- (8,4) -- (4,4) -- cycle;
       \draw (6,2) node[black, scale=1, rectangle split, rectangle split parts=2, draw,  fill=white]{
        Phase III
        \nodepart{second}
        BER=1/2
      };    

      \draw (0,0) -- (4,4);
      \draw (1,2) node[scale=0.6] {$\log k=\alpha \log n$};
      \draw (0.9,1.9) edge[->] (1.5, 1.5);

    \end{tikzpicture}
    \caption{A schematic view of the main results. The three  shaded regions correspond to the three different parts of the theorem. Only the asymptotic behaviour is presented in the figure.}
    \label{fig:01}
  \end{center}
\end{figure}
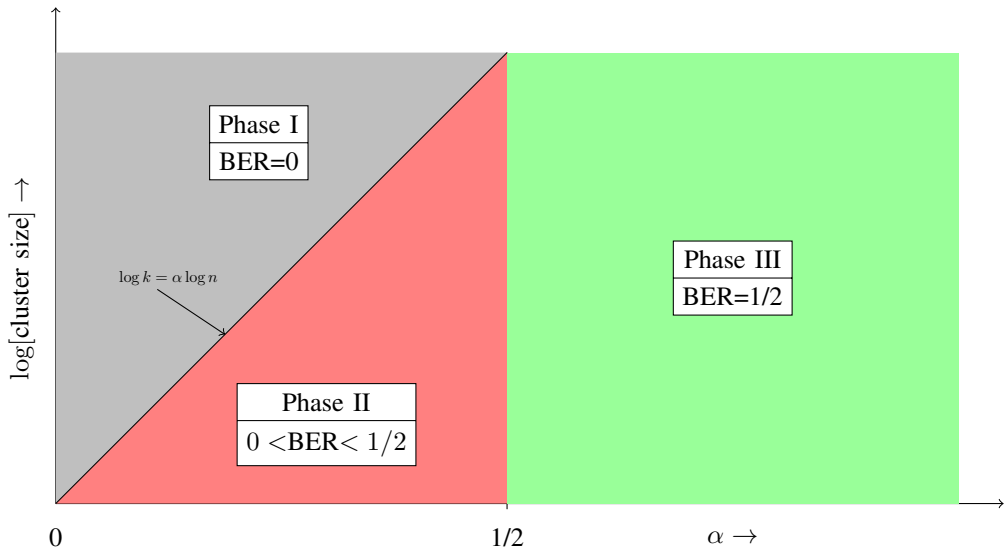


To give an outline of our results, suppose that the rating matrix is of size $n \times n$ and the erasure probability $\epsilon = 1-c/n^{\alpha}$ for $c >0$, $\alpha \in [0,1]$ and the BSC error probability $p$.  We note that $\alpha$ controls the rate at which the erasure probability approaches 1. This rate plays a crucial role in determining the performance of \texttt{PAF}. Suppose the rows, as well as the columns are clustered with  each cluster having size $k$.  We identify three different performance regimes, which are illustrated in Fig.\ \ref{fig:01}, in the limit as $n \to \infty$. 
\begin{itemize}
\item When $\alpha \in [0,1/2)$, if the cluster size ($k$) is greater than $n^{\alpha -\gamma_n}$ where $\gamma_n \rightarrow 0$, then the BER approaches 0 (Phase I of Fig.\ \ref{fig:01}). This result in stated in Theorem \ref{thm:01} of Section \ref{subsec:main}.
\item When $\alpha \in [0,1/2)$, if the cluster size ($k$) is less than $n^{\alpha -\gamma}$, $\gamma >0$, the BER is bounded away from zero and a lower bound is  obtained in terms of the BSC error probability and $\gamma$ (Phase II of Fig.\ \ref{fig:01}). This result is stated in Theorem \ref{thm:02} of Section \ref{subsec:main}. Further, in Theorem \ref{thm:02}, we also identify the exact limiting BER (except for some special cases of $\gamma$) and also the optimal parameter for  \texttt{PAF}.
\item  For $\alpha >1/2$, the BER always approaches 1/2 (Phase III of Fig.\ \ref{fig:01}). This result is stated in Theorem \ref{thm:03} of Section \ref{subsec:main}.
\item We then study a lower bound on the performance  of such a  recommender, and compare this with the performance of \texttt{PAF}. We state this result in Theorem \ref{thm:04} of Section \ref{subsec:main}.
\end{itemize}
The main results are proven in Section \ref{sec:proof:01}, Section \ref{sec:proof:02} and Section \ref{sec:proof:03}, followed by a conclusion in Section \ref{sec:con}. We present the proofs of several related lemmas in the Appendix.

\section{The Algorithm and its Performance on Real Data}
\label{sec:algo}
In Section \ref{subsec:locPop}, we describe the $\mathtt{PAF}$ algorithm, and in Section \ref{subsec:exp} we evaluate its performance on some real datasets.

\subsection{The \texttt{PAF} algorithm}
\label{subsec:locPop}

Suppose $\mathbf Y$ is an $m\times n$  user-item matrix with entries in $\{0,1,*\}$. The rows represent the users and the columns represents the items. If the $(i,j)$th entry $\mathbf Y(i,j)$ is 1 (or 0), then we interpret it as ``user $i$ likes (or does not like) the item $j$''. A `$*$' indicates an unobserved rating. Upon observing $\mathbf Y$, we want to recommend an item (a column) to user 1.  For rows $i$ and $j$, consider the number of entries  that they agree on:
\begin{equation}
\label{eq:def:01}s_{ij}:=\sum_{k=1}^n\mathbf 1_{\{\mathbf Y(i,k)\neq *\}}\cdot \mathbf 1_{\{\mathbf Y(j,k)\neq *\}}\cdot\mathbf  1_{\{\mathbf Y(i,k)=\mathbf Y(j,k)\}},
\end{equation}
where $\mathbf 1_{\{.\}}$ denotes the indicator function.
We use the following $\mathtt{PAF}$ algorithm to recommend an item $j_0$ to user 1.

\vspace{0.03in}
\noindent
\begin{center}
\fbox{\parbox{6in}{ 
$\mathtt{PAF}(T):$
\begin{enumerate}
\item ({\bf Select the top $T$ nearest rows}) Compute $s_{1i}$, for 
      $i=1,2,3, \dots,m$. Select the top $T$ rows with the highest values of similarity, where $T$ is a parameter whose choice is discussed later. 
      
\item ({\bf Pick the most popular column}) Amongst the columns $j$ such that $\mathbf Y(1,j)=*$, select the column having maximum number of 1's amongst the top $T$ neighbors. Break ties randomly. 
\end{enumerate}
}}
\end{center}
Suppose we represent each row by a vertex in a graph with an edge between vertex $i$ and $j$ iff $s_{ij}>0$. Then to recommend an item to user 1, the above algorithm depends only on the rows neighboring to user 1, and chooses the most popular item amongst the top few neighbors. Let $\bar d$ denote the average degree of a vertex in this graph. Then the complexity of Step 1 is $O(\bar d m)$, and since $\bar d$ is usually much smaller than $m$, the overall complexity of Step 1 is {\em low}.

We note that several variants of the similarity metric are feasible, but as we show below, the \texttt{PAF} algorithm described above has competitive performance on real datasets, and is also amenable to analysis. 

\subsection{Experimental results and discussion}
\label{subsec:exp}
We consider the MovieLens data \cite{MovieLens} (consisting of 1,000,209 ratings for 3952 movies made by 6040 users) as well as a snapshot of the Netflix data \cite{netflixprize} (consisting of 818,229 ratings for 4289 movies made by 7457 users, obtained in year 2000). For both MovieLens and Netflix,  the ratings are integers between 1 and 5. To apply the \texttt{PAF} algorithm, we quantize the ratings: 4 and 5 are mapped to 1 (``recommended'' movies), while 1, 2 and 3 are mapped to 0 (``not recommended'' movies). 
We split the ratings as train and test data as follows. 
For each user, we randomly hide $30\%$ of the ratings, and use these as the test data. We train our algorithms on the remaining data.  We can check correctness of a recommendation only if the rating of the recommended movie is hidden.

\begin{figure}
\centering
\includegraphics[width=3.5in]{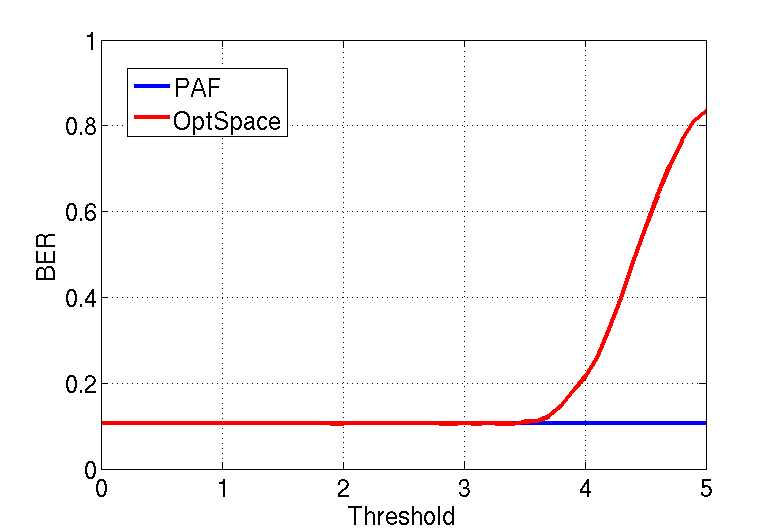}
\caption{Performance comparison of OptSpace with \texttt{PAF} for the MovieLens dataset (1,000,209 ratings), as the threshold used to quantize the estimated values of OptSpace changes.}
\label{fig:varying:thr}
\end{figure}

We compare the performance of the $\mathtt{PAF}$ algorithm with OptSpace (the algorithm proposed in \cite{Montanari1}). OptSpace uses ratings on the scale 1-5 as input and outputs real valued rating estimates. Since OptSpace outputs real values, in order to compute the BER, we map the predicted ratings below 3.5 to 0, and the predicted ratings above 3.5 to 1. The BER is computed over the same set as for the $\mathtt{PAF}$ algorithm. 
Using a threshold of 3.5 is not necessarily optimal. In Fig.\ \ref{fig:varying:thr}, we see how the performance of OptSpace vary for the MovieLens dataset as we change the threshold from 1 to 5. When the threshold is 0, OptSpace estimate all the entries as 1's, and it's performance exactly matches with \texttt{PAF}. At the other extreme, when the threshold is 5, OptSpace estimates everything as 0's, and it's performance degrades. Because of the rating quantization scheme that we use (mapping $\{1,2,3\}$ to 0, and $\{4,5\}$ to 1), only a threshold between 3 and 4 makes sense. Since we do not see any significant improvement of performance by optimizing over this threshold, we continue to use 3.5 as the threshold. Similar behavior is also observed  for the Netflix dataset.
For both \texttt{PAF} and OptSpace, we have chosen the parameters that yield the best performance on the test data.

\begin{table}[!h]
  \centering
 \caption{Comparison of BER and RMSE  of \texttt{PAF} with OptSpace}
\subtable[Original MovieLens data (1,000,209 ratings)]{
\centering
  \begin{tabular}{|c|c|c|}    
    \hline
   &  $\mathtt{PAF(100)}$ & OptSpace\\
    \hline
    BER & 0.103 & 0.108\\
\hline
RMSE &  0.748 & 0.733 \\
     \hline
  \end{tabular}
\label{tab:ber:a}
}\hspace{0.2in}
\subtable[A snapshot of Netflix data (818,229 ratings)]{
\centering
  \begin{tabular}{|c|c|c|}    
    \hline
    & $\mathtt{PAF(80)}$ & OptSpace\\
    \hline
    BER & 0.116 & 0.127\\
    \hline
RMSE & 0.942 & 0.742 \\
     \hline
  \end{tabular}
\label{tab:ber:b}
}\hspace{0.2in}
\subtable[MovieLens data, after removing the popular movies (1,000,209 ratings)]{
\centering
  \begin{tabular}{|c|c|c|}    
    \hline
    & $\mathtt{PAF(55)}$ & OptSpace\\
    \hline
    BER & 0.321 & 0.327\\
     \hline
     RMSE & 1.010 & 0.901 \\
\hline
  \end{tabular}
  \label{tab:ber:c}
}
 \label{tab:ber}
\end{table}

 Table \ref{tab:ber:a} and \ref{tab:ber:b} show that in terms of BER, the $\mathtt{PAF}$ algorithm and OptSpace are close for both the MovieLens as well as the Netflix data.
 We see that \texttt{PAF} is comparable to OptSpace. We also compare both these methods in terms of their root mean square error (RMSE). To compute the RMSE for \texttt{PAF} we map the binary estimates to a scale a 1-5 as the following. A 0 is mapped to 2 (average of $\{1,2,3\}$), and a 1 is mapped to 4.5 (average of $\{4,5\}$). (Although this mapping is not necessarily optimal, we do not try to optimize it.) From the RMSE values in Table \ref{tab:ber:a} and Table \ref{tab:ber:b}, we see that for the MovieLens dataset both the algorithms are comparable and for the snapshot of Netflix dataset, OptSpace performs better than \texttt{PAF} in terms of RMSE. A comparison of  this with the BER comparison  tells us that improvement in RMSE has little impact on BER, which is a reflection of the poor confidence interval in the estimate. For this reason, we believe binary alphabet and the BER metric are more relevant for these datasets. This point is discussed further in Section \ref{subsec:model} in the paragraph {\bf Why binary}.

\begin{figure}
\centering
\subfigure[]{\includegraphics[scale=0.35]{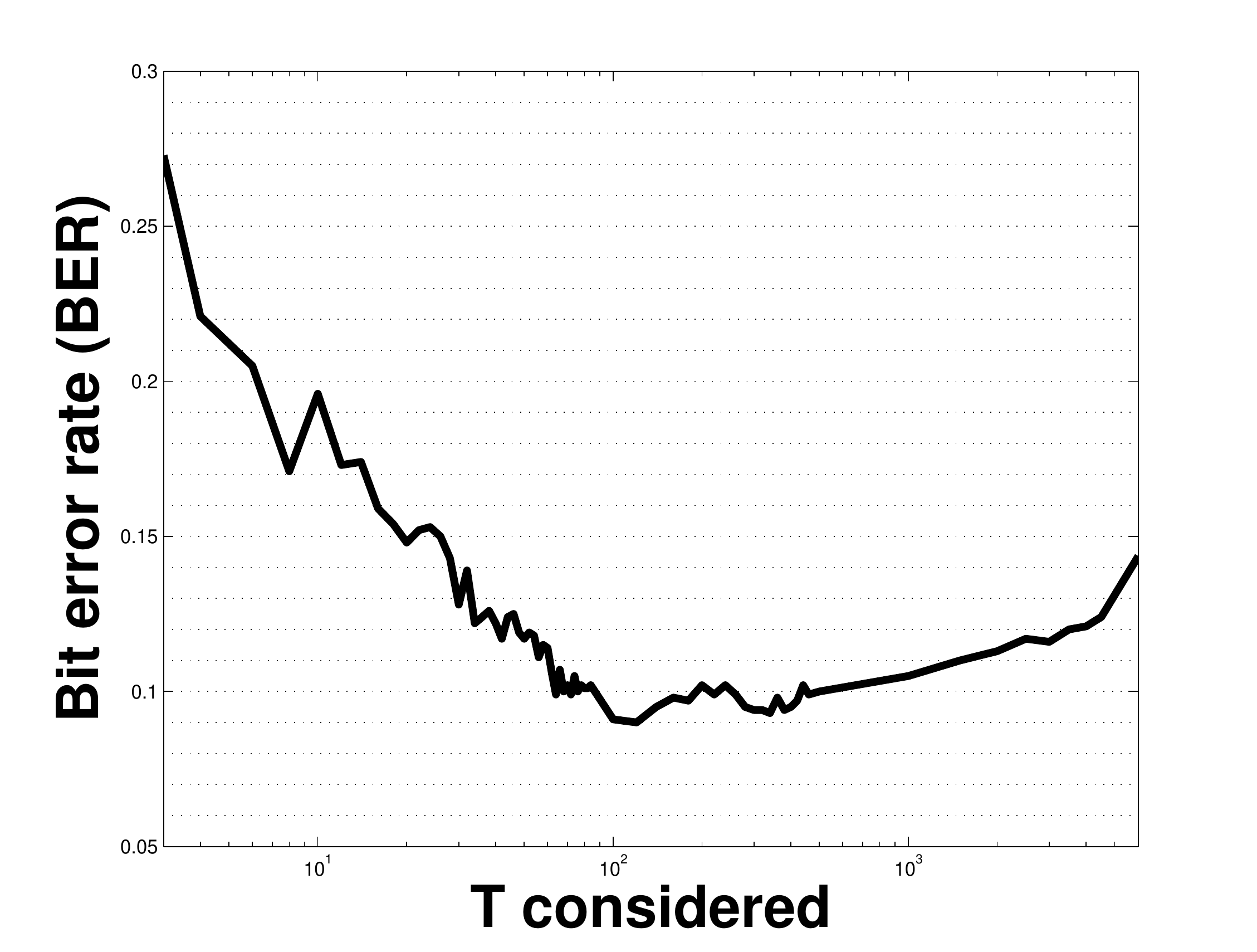}
\label{fig:comp:a}
}
\subfigure[]{\includegraphics[scale=0.35]{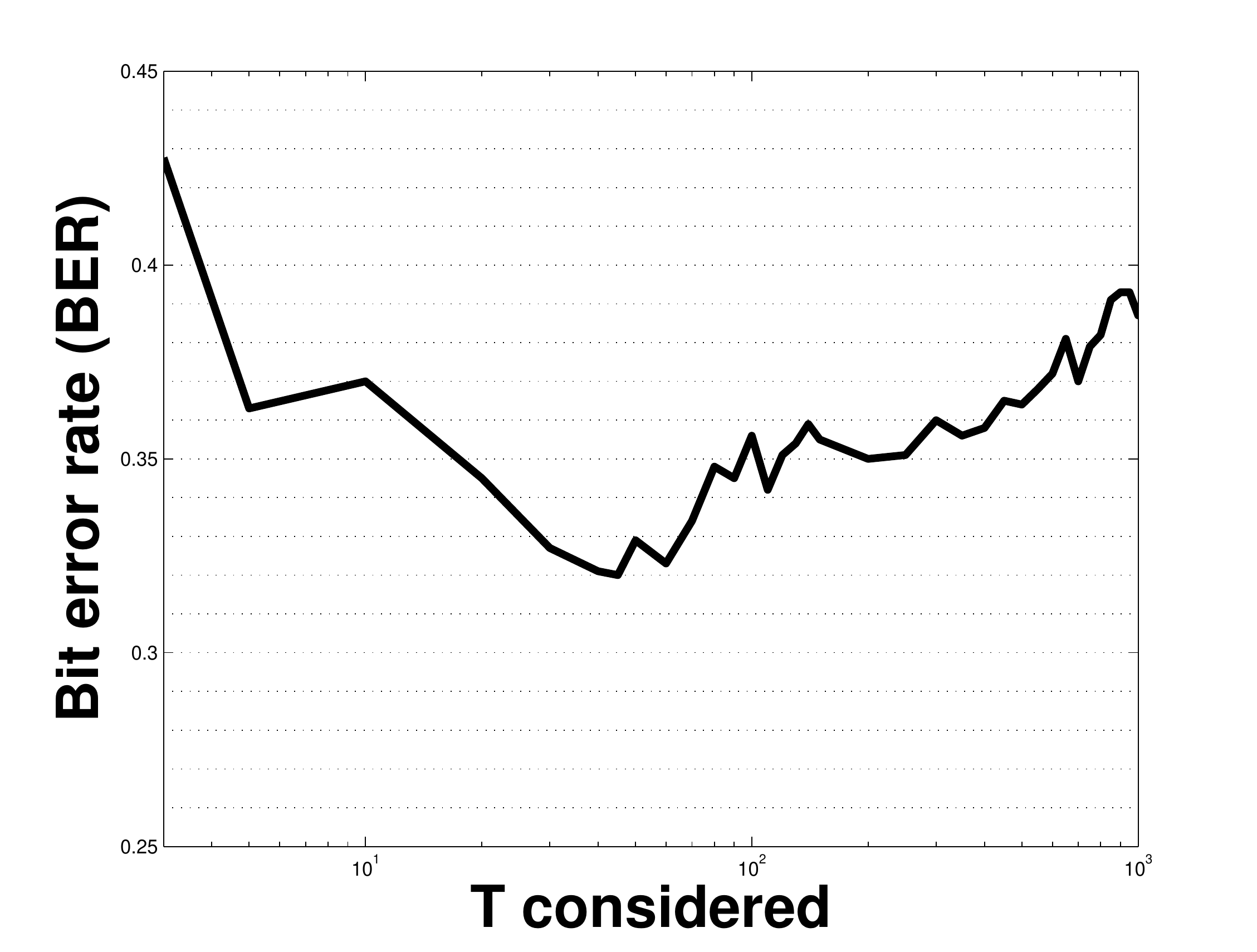}
\label{fig:comp:b}
}
\caption{Bit error rate of \texttt{PAF} for different values of $T$, for the MovieLens data with  1,000,209 ratings. While (a) compares the BER for the original MovieLens data, (b) compares the BER for the MovieLens data after filtering out the popular movies with  more than 60 \% of their ratings as 1's. }
\label{fig:comp}
\end{figure}

 Fig.\ \ref{fig:comp:a} shows how  the $\mathtt{PAF}(T)$ performs for different values of $T$ for the MovieLens data. We see that the BER is minimized around $T=100$. We also note that for the snapshot of Netflix data we consider, the BER is minimized at around $T=80$. In Theorem \ref{thm:02} of Section \ref{subsec:main}, we show that the minimum BER is achieved at $T=k$ (the ``true'' cluster size), and hence the minimum in Fig.\  \ref{fig:comp:a} is related to the degrees of freedom in the data.

If we use $T=m$, then we get the global popularity algorithm, and it has a BER of about  0.16 for the MovieLens dataset.  This indicates that the dataset has several movies, which are popular amongst most users, and hence their ratings are easy to predict. The true test of a collaborative filter is on datasets where a single row or column does not reveal too much information about its missing entries. Since $\mathtt{PAF}$ algorithm is biased towards globally popular movies, to test its performance further, for the MovieLens dataset we remove all movies with more than $60\%$ ratings as 1. 
Even for this ``filtered'' dataset, we see from Table \ref{tab:ber:c} that the  $\mathtt{PAF}$ algorithm and OptSpace are comparable. Fig.\ \ref{fig:comp:b} shows that the minimum BER is achieved when $T$ is around 55.

\begin{remark}
If we look at \texttt{PAF}, we see that most of its computational time is spent in finding the row correlations. As the data evolves with time, in the sense that new user/movie enters in the data or users rate more existing movies, then the row correlations can be updated efficiently since usually only a few of the row correlations are affected at a time. 
\end{remark}

In summary, the \texttt{PAF} algorithm yields competitive performance on real data, even though it used only quantized ratings (as against to 1-5 for OptSpace).
To explain the competitive performance of the $\mathtt{PAF}$ algorithm, in the following section,  we analyze its performance for a binary matrix model introduced in \cite{Aditya1}.

\section{Analysis of the $\mathtt{PAF}$ Algorithm}
In Section \ref{subsec:model} we describe our mathematical model (first introduced in \cite{Aditya1}, \cite{Aditya2}) and in Section \ref{subsec:main} we state and discuss our main results. But before we begin with analyzing \texttt{PAF}, we set up some notation.

\noindent
{\bf Notation:}
By $X\sim B(n,p)$ we mean that a random variable $X$ is binomially distributed with parameters $n$ and $p$. 
For two real valued functions $f(n)$ and $g(n)$, if there exist strictly  positive $M$ and $n_0$ such that $|f(n)| \le M |g(n)|$ for all $n > n_0$, then we denote $f(n)=O(g(n))$ and $g(n)=\Omega(f(n))$. If $f(n)=O(g(n))$ and $f(n)=\Omega(g(n))$ then we say $f(n)=\Theta(g(n))$. We say $f(n)=o(g(n))$ if $\lim_{n\rightarrow \infty} \frac{f(n)}{g(n)}=0$, and  
$f(n) \doteq g(n)$ if $\lim_{n\rightarrow \infty} \frac{f(n)}{g(n)}=1$. For a sequence of real valued functions $\{f_i(n)\}_{i\in I}$ and $g(n)$, if there exist strictly positive  $M$ and $n_0$ (both independent of $i$) such that for $i\in I$ and for $n >n_0$ we have $|f_i(n)| \le M|g(n)|$, then we denote $\{f_i(n)\}_{i\in I}=O(g(n))$. Other order notations for sequence of functions are defined in a similar manner. For a matrix $\mathbf X$, $\mathbf X(:,j)$ denotes the $j$th column of $\mathbf X$. For a vector $\bar y\in \{0,1,*\}^n$ where $*$ denotes an erasure, $|\bar y|_0$, $|\bar y|_1$ and $|\bar y|$ represent number of 0's, number of 1's and the total number of 0's and 1's respectively. For a sequence of events $\{A_n\}$, if $P[A_n]\rightarrow 1$ with $n$, then we say that  $A_n$ occurs w.h.p. . For parameters that depend on the data size $n$ (e.g., $\epsilon$, $k$, etc.), we do not show this dependence explicitly unless it is not clear from the context.

\label{sec:analysis}
\subsection{The Data Model}
\label{subsec:model}
\begin{figure}
    \begin{center}
      \begin{tikzpicture}
      \draw (-0.3, 2) node[black, scale=1, rectangle,  draw,  fill=white]
      {$ \underbrace{\mathbf X}_{\text{\parbox{3cm}{Matrix with unknown \\ row and column clusters}}}  \xrightarrow{\hspace{.5cm} BSC(p) \hspace{.5cm}}  \underbrace{\mathbf X_e}_{\text{Matrix with errors}}  \xrightarrow{\hspace{.4cm} Erasure(\epsilon) \hspace{.4cm}} \underbrace{\mathbf Y}_{\text{\parbox{3cm}{The observed matrix with\\ errors and erasures}}}$};
    \end{tikzpicture}
  \end{center}
\caption{Summary of the data model.}
\label{fig:flow}
\end{figure}
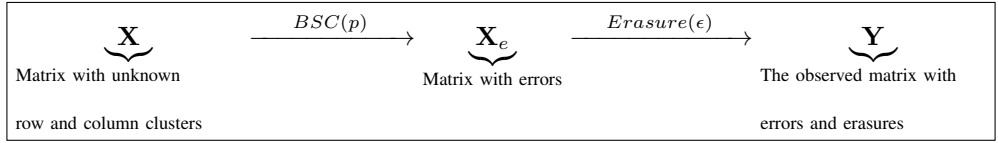

We consider an $n\times n$ matrix $\mathbf X$ whose entries are binary. The rows of the matrix represent users and the columns represent items. Suppose $\mathcal A=\{A_i\}_{i=1}^r$ and $\mathcal B=\{B_i\}_{i=1}^r$ are two partitions of $[1:n]$, representing sets of similar users and items.  We call the sets $A_i\times B_j$ clusters, and call $A_i$'s ($B_j$'s) the row (column) clusters.  We assume that for all $i=1,2,...,r$, we have $|A_i|=|B_i|=k$.  The matrix $\mathbf X$ is constant over the cluster $A_i\times B_j$ and the entries are i.i.d.\ Bernoulli (1/2) \footnote{A  random variable $X$ is called Bernoulli($p$), if $Pr[X=1]=p$, and $Pr[X=0]=1-p$.} across the clusters. Formally, if $(p,q)\in A_i\times B_j$, then $\mathbf X(p,q)=\chi_{ij}$ where $\{\chi_{ij}\}_{i,j=1}^r$ are i.i.d. Bernoulli(1/2). 
The observed matrix $\mathbf Y$ is obtained by passing the entries of $\mathbf X$  independently through  binary  symmetric channel (BSC) (defined below) with parameter $p$, and then through a binary erasure channel (defined below) with erasure probability $\epsilon$. The entries of the  observed matrix $\mathbf Y$ are from $\{0,1,*\}$, where $*$ denotes an erased entry.  Fig.\ \ref{fig:flow} Summarizes our data model.

The BSC is a  binary input, binary output channel that makes an error with probability $p$ (\cite{Cover1}). In our case, it models noisy behavior of users. In the binary erasure channel, every bit is erased with probability $\epsilon$, and the receiver knows which bits have been erased (\cite{Cover1}).
The erasure channel models  the missing entries in the rating matrix.

{\bf Why binary?:} We consider the case of binary entries for simplicity, and like in \cite{Aditya2}, this can be relaxed to allow any finite alphabet. The choice of the binary alphabet not only leads to a simpler description of the main ideas, but as explained below, it is also a case of practical interest.
\begin{itemize}
\item For datasets such as Netflix, even the best known methods have a  root mean square error (RMSE) of 0.8567 \cite{netflix-leaderboard}, which on a scale of 1-5 elicits poor confidence in the estimate. This is because, even in the absence of variance (i.e., when all the contribution to RMSE comes from the bias), the confidence interval for such an estimate is $\pm 0.8567$, which shows poor confidence on a scale of 1-5. 
 However, the task of determining whether a movie is liked (say rating $\geq 4$) or not can be done with more reliability, suggesting the importance of the binary alphabet in what appears to be very noisy data. (In fact, in Section \ref{subsec:exp}, we saw that the \texttt{PAF} algorithm uses quantized inputs on the binary scale (instead of 1-5) but still yields competitive performance compared to OptSpace, which uses the unquantized inputs.)
\item In many datasets, users tend to rate items either very high or very low. For example, this was observed in a recent study by Youtube \cite{youtube-binary}, \cite{youtube-binary-1}, which prompted the switch to a binary rating scale instead of 1-5.
\end{itemize}

We also note that all our results can be extended to the case when $\mathbf X$ is $m
\times n$ and the clusters are nonuniform, provided $m=\Theta(n)$ and all the cluster sizes are of same order. Since the non-uniform case does not offer any additional new insights, in this paper we have chosen to use the uniform case, which leads substantially simpler notation.

\subsection{Main Results and Discussion}
\label{subsec:main}
Upon observing $\mathbf Y$, suppose $\mathtt{PAF}(T)$ recommends a column $j_{max}$. The probability of error for this recommendation is 
$$P_e[\mathtt{PAF}(T)]=Pr[\mathbf X(1,j_{max})=0].$$
Here we study how the  $\mathtt{PAF}$ algorithm performs for the  matrix model discussed above, and identify three different performance regimes based on the erasure rate and the cluster size. In the following, we assume that the erasure probability $\epsilon=1-\frac{c}{n^\alpha}$ for some $c>0$ and $\alpha>0$, and assume that the true cluster size $k$ is known. The value of $\alpha$ determines the rate at which the erasure probability approaches unity as $n$ grows. We have the following theorems.

\subsubsection{Low Erasure Rate, Large Cluster Size}

This regime is illustrated by the Phase I of  Fig.\ \ref{fig:01} and the main result is as follows. Recall that without loss of generality, we recommend an item to user 1.


\begin{theorem}[{\bf $\alpha <1/2$, large cluster size}]
  \label{thm:01}
Assume that $\alpha \in (0,1/2)$,   and the BSC error probability  $p\in [0,1/2)$. 
Suppose there exists a sequence $\gamma_n\ge 0$ such that $\gamma_n \rightarrow 0$ and $k \ge n^{\alpha-\gamma_n}$. 
Then the following are true.

a) If $k=o(n)$, then $P_e[\mathtt{PAF}(k)]\rightarrow 0$.

b) If $k=\Theta(n)$ , then $P_e\left[\mathtt{PAF}(k)\big|\text{ not all entries of the 1st row of $\mathbf X$ are 0's}\right]\rightarrow 0$. 

 For $\alpha =0$, the error probability goes to zero as long as $k$ increases to infinity with $n$.
\end{theorem}

This result is proved in Section \ref{sec:proof:01} but next we describe the main intuition behind the result.
When $\alpha <1/2$, there are enough  samples to distinguish the neighbors from $A_1$ (``good'' neighbors) from the neighbors outside $A_1$ (``bad'' neighbors). In fact, all the top $k$ neighbors selected by the $\mathtt{PAF}$ algorithm are good with high probability.
Moreover, when  $\gamma_n \to 0$, we show that the most popular column has overwhelming number of 1's compared to 0's. We then show that this cannot happen unless the true rating of the most popular column is 1 with high probability (w.h.p.).

\begin{remark}
When $r$ is bounded (i.e., $k=\Theta(n)$), we need the assumption that not all entries in the 1st row of $\mathbf X$ are 0's, because  there is a nonzero probability that all entries of the 1st row of $\mathbf X$ are 0's.  In this case we will always make a wrong recommendation. 
\end{remark}

\begin{remark}
It is also of interest to know the rate at which the error probability goes to zero. The convergence rate crucially depends on $\gamma_n$ in a non-trivial manner and we are unable to find a clean bound. However, for  $\gamma_n=0$ we can find a bound on the error probability, and we have  $P_e[\mathtt{PAF}(k)] = O\left(1/c_1^{\sqrt{\log n}}\right)$ for some $c_1>1$. We also note that this bound is not tight in general.
\end{remark}

\subsubsection{Low Erasure Rate, Small Cluster Size}
From the empirical results in Section \ref{subsec:exp}, we see that $0<\text{BER}<1/2$. If we assume that our asymptotic model is applicable to the data size considered, then the regime of Theorem \ref{thm:01} does not seem to capture this. Theorem \ref{thm:02} stated below identifies a regime where the asymptotic BER of the \texttt{PAF} algorithm is bounded away from both 0 and 1/2. (Phase II of Fig.\ \ref{fig:01} illustrates this regime.) 

\begin{theorem}[{\bf $\alpha <1/2$, small cluster size}]
\label{thm:02}
Assume that $\alpha \in (0,1/2)$,   and the BSC error probability  $p\in [0,1/2)$.
Suppose there is a constant $\gamma \in ( 0, \alpha]$ and $g_n=o(1)$ such that the cluster size $k= n^{\alpha -\gamma+g_n}$. Then the limit $\lim_{n\rightarrow \infty} P_e[\mathtt{PAF}(k)]$ exists, and we have the following.
\begin{itemize}
\item If $1/\gamma$ is not an integer, then  $$\lim_{n\rightarrow \infty} P_e[\mathtt{PAF}(k)]  = \frac{p^{\left\lfloor \frac{1}{\gamma}\right\rfloor}}{p^{\left\lfloor \frac{1}{\gamma}\right\rfloor} + (1-p)^{\left\lfloor \frac{1}{\gamma}\right\rfloor}}.$$
\item If $1/\gamma$ is an integer, then 
$$\frac{p^{\frac{1}{\gamma}}}{p^{\frac{1}{\gamma}} + (1-p)^{ \frac{1}{\gamma}}} 
\le \lim_{n\rightarrow \infty} P_e[\mathtt{PAF}(k)] 
\le \frac{p^{\frac{1}{\gamma}-1}}{p^{\frac{1}{\gamma}-1} + (1-p)^{ \frac{1}{\gamma}-1}}.$$
\end{itemize}
Moreover $T=k$ is optimal, in the sense, that $\forall T$,
$$\lim_{n\rightarrow \infty} P_e[\mathtt{PAF}(k)] \le \lim\inf_{n\rightarrow \infty} P_e[\mathtt{PAF}(T)]. $$
\end{theorem}
We prove this theorem in Section \ref{sec:proof:02}, but below we provide some intuition. 

As in Theorem \ref{thm:01}, when $\alpha < 1/2$, for $T = k$ most neighbors picked are good with high probability. However, since $\gamma > 0$, the number of 1's for the most popular movie is concentrated on $\lfloor 1/\gamma \rfloor$ when $1/\gamma$ is not an integer (and is concentrated on $\{1/\gamma-1, 1/\gamma \}$ when $1/\gamma$ is an integer), which is finite. Thus, even though the algorithm picks the good neighbors, it fails to average out the noise in the ratings completely, leading to a BER bounded away from 0. 

Furthermore, Theorem \ref{thm:02} states that in the limit as $n \to \infty$, $T=k$ is optimal. This is expected since for $T < k$ we do not use the full set of good neighbors, and for $T > k$, we pick bad neighbors. 
As $T$ approaches $n$, the \texttt{PAF} algorithm approaches the global popularity algorithm, and for our mathematical model, its BER is 1/2. We note that for the MovieLens dataset with popular movies removed, Fig.\ \ref{fig:comp} suggests an optimal value of $T=55$, which is a reflection of the user cluster size.

\subsubsection{High Erasure Rate}
The above two theorems discuss the case when $\alpha <1/2$. In this case, w.h.p. the $\mathtt{PAF}$ algorithm can filter out the bad neighbors. But when $\alpha >1/2$, there are few samples to distinguish the good neighbors from the bad ones. In fact, amongst the top $T$ neighbors, only a vanishingly small fraction are good neighbors. This forces the BER to approach 1/2, and is stated in Theorem \ref{thm:03} below, which is proved in Section \ref{sec:proof:03}.
\begin{theorem}[$\alpha >1/2$]
  \label{thm:03}
Assume that  $\alpha > 1/2$,  the BSC error probability  $p=0$, and $k=o(n)$. 
Then $\forall T$, 
$$P_e[\mathtt{PAF}(T)]\rightarrow 1/2.$$
\end{theorem}

In the regime of Theorem \ref{thm:03}, the errors occur mainly due to the fact that the $\mathtt{PAF}$ algorithm cannot identify the good neighbors. Some  side information about the similarity amongst users (for example information about social connections, locations, etc.) would help the algorithm to find the good neighbors. In Fig.\ \ref{fig:01}, Phase III represents this high erasure rate regime.

\begin{remark}
  In the regime of Theorem \ref{thm:03}, i.e., for $\alpha >1/2$, we need that $r\rightarrow \infty$ to prove that BER goes to 1/2. If $r$ stays bounded, then we believe that the BER would be bounded away from 1/2 and 0.  But we are unable to prove this yet.
\end{remark}

\begin{figure}
\centering
\includegraphics[width=3.0in]{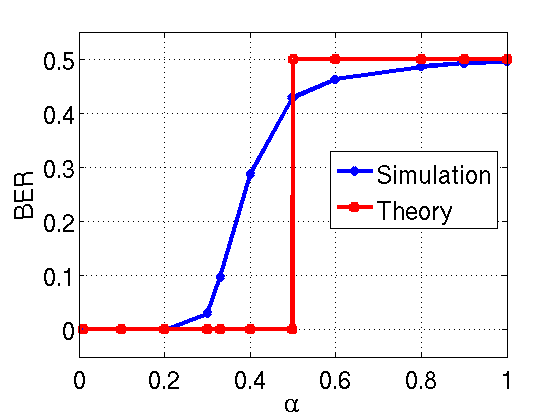}
\caption{Comparison of the asymptotic analysis vs simulation for $n=1000$, $k=10$, $p=0.2$, and varying $\alpha$.}
\label{fig:simulation}
\end{figure}
{\bf A numerical example:} Given the above three theorems describing the asymptotic performance of $\mathtt{PAF}$ under various regimes, it is of interest to understand if such asymptotics are valid for finite data size. To answer this, we simulate  datasets using the our data model with $n=1000$, $k=10$, $p=0.2$, and with varying $\alpha$. Fig.\ \ref{fig:simulation} shows that even for this small dataset, the asymptotic theory matches well with the simulation for $\alpha <1/3$ and $\alpha >1/2$.   Since $k=n^{1/3}$, $\alpha <1/3$ represents the regime of Theorem \ref{thm:01}. Similarly, $\alpha >1/2$ represents the regime of Theorem \ref{thm:03}. In the regime of Theorem \ref{thm:02} (i.e., for $1/3 <\alpha <1/2$), there is a gap between the asymptote and simulation, and we need to consider larger dataset to reduce this gap.

\subsubsection{Suboptimality of \texttt{PAF}}
Having seen the performance of \texttt{PAF} in the above theorems, from a mathematical perspective it is natural to ask  if \texttt{PAF} is optimal  for the above data model.    Let $P_{e}(n)$ denotes the error probability of a given recommender, parametrized by the matrix size $n$. 

\begin{theorem}
  \label{thm:04}
Suppose the BSC error probability  $p\in [0,1/2)$. 
\begin{itemize}
\item {\bf Converse:}
If $k^2\le n^{\alpha-\gamma +g_n}$ for $\gamma\in [ 0, \min (\alpha,1)]$ and $g_n=o(1)$, then for any recommender
$$\lim\inf_{n\rightarrow \infty} P_{e}(n) \ge \frac{p^{\left\lfloor \frac{1}{\gamma}\right\rfloor}}{p^{\left\lfloor \frac{1}{\gamma}\right\rfloor} + (1-p)^{\left\lfloor \frac{1}{\gamma}\right\rfloor}}.$$

\item {\bf Achievability:}
Assume $\alpha \in (0,1/2)$, and suppose that $k=o(n)$.
If there exists $\gamma_n=o(1)$ such that $k^2\ge n^{\alpha -\gamma_n}$, then there exists an algorithm (described in the proof) s.t.,
$$P_{e}(n) \rightarrow 0.$$ 
Moreover, if $k^2 = n^{\alpha-\gamma +g_n}$ for $\gamma\in ( 0, \min (\alpha,1)]$ and $g_n=o(1)$, then
$$ \frac{p^{\left\lfloor \frac{1}{\gamma}\right\rfloor}}{p^{\left\lfloor \frac{1}{\gamma}\right\rfloor} + (1-p)^{\left\lfloor \frac{1}{\gamma}\right\rfloor}} 
\le \lim_{n\rightarrow \infty} P_{e}(n) \le 
\frac{p^{\left\lceil \frac{1}{\gamma} -1\right\rceil}}{p^{\left\lceil \frac{1}{\gamma}-1\right\rceil} + (1-p)^{\left\lceil \frac{1}{\gamma}-1 \right\rceil}}.$$
\end{itemize}
\end{theorem}

\begin{remark}
We note that the lower and upper bound in the final expression of Theorem \ref{thm:04} are identical, unless $1/\gamma$ is an integer.
\end{remark}
The  lower bound in the converse is obtained by using an oracle, which tells us the true clusters ($\mathcal A$ and $\mathcal B$), and  then using techniques similar to ones used in proving  Theorem \ref{thm:02}. The achievability proof uses that for $\alpha<1/2$ and $r>c_1\log n$, w.h.p.\ we can cluster the matrix correctly. Then  the result for $k^2\ge n^{\alpha -\gamma_n}$ follows from arguments similar to those used in proving Theorem \ref{thm:01}; and the result for $k^2 = n^{\alpha-\gamma +g_n}$ is obtained by using arguments similar to those used in proving Theorem \ref{thm:02}. A more detailed proof is presented in Section \ref{sec:proof:04}.

Comparing Theorem \ref{thm:01} and Theorem \ref{thm:02} with Theorem \ref{thm:04}, we see that \texttt{PAF} is suboptimal. But  \texttt{PAF} is computationally faster than the algorithm that achieves the bounds in Theorem \ref{thm:04} (described in the proof), since it does not require to do any explicit clustering of the rows and the columns. This is one of the main reasons why we consider \texttt{PAF} in this paper (instead of the clustering based algorithm in \cite{Aditya2} or in the proof of Theorem \ref{thm:04}). In Section \ref{subsec:exp}, we have already seen the competitive performance of \texttt{PAF} on real world datasets, which makes  \texttt{PAF} even more appealing.

In the following, we present the proofs of these four theorems.

\section{Proof of  Theorem \ref{thm:01}}
\label{sec:proof:01}

The  $\mathtt{PAF}$ algorithm has two steps. First we find the neighbors, and then we recommend using the popularity amongst the neighbors. We analyze the errors in these steps separately.

\subsection{Analysis of Step 1 of the Algorithm}
\label{subsec:step1}

We show that for $\alpha <1/2$,  w.h.p.\ the top $k$ rows are all from the cluster of user 1, namely $A_1$ \footnote{If for a row cluster $A_i$, $\mathbf X_{A_i}$ ($\mathbf X$ restricted to the rows of $A_i$) is identical to $\mathbf X_{A_1}$, then for all practical purpose we can include all the rows of $A_i$ in $A_1$ itself. Throughout this proof, we assume that the rows from all the clusters identical to $A_1$ have already been included in $A_1$. Thus for $i\not \in A_1$, the $i$th row and the 1st row differ at least at one column cluster.}. First we obtain the following two lemmas that will help us in proving this. Recall that $p$ denotes the error probability of the BSC.
\begin{lemma}[{Overlap with rows within cluster}]
  \label{lemma:01}
For any $\delta\in (0,1)$, we have  w.h.p. for all $i$ in $A_1$, 
\[s_{1i} \ge n^{1-2\alpha}((1-p)^2 + p^2)(1-\delta).\]
\end{lemma}
\begin{proof}
We see that $s_{11}\sim B(n, 1-\epsilon)$ and for $i\in A_1\backslash\{1\}$, $s_{1i}\sim B(n, (1-\epsilon)^2((1-p)^2 + p^2))$. In other words, for $i\in A_1\backslash\{1\}$, $s_{1i}$ is a Binomial random variable with $\mathbb E[s_{1i}]= n(1-\epsilon)^2((1-p)^2 + p^2)= cn^{1-2\alpha}((1-p)^2 + p^2)$; and $s_{11}$ is a Binomial random variable with $\mathbb E[s_{11}]=n(1-\epsilon)= cn^{1-\alpha}$. 
The lemma is now a direct consequence of the Chernoff bound \cite[Theorem 1.1]{Dubhashi1}, together with the union bound.
\end{proof}

\begin{lemma}[{Overlap with rows outside cluster}]
  \label{lemma:02}
There is a constant $c_1\in (0,1)$, such that for 
any $\delta\in (0,1)$, we have w.h.p. for all $i$ outside $A_1$,
\[s_{1i}\le n^{1-2\alpha} [(1-p)^2+p^2 - c_1(1-2p)^2](1+\delta).\]
\end{lemma}
\begin{proof}
The proof is  given in Appendix \ref{proof:lemma:02}.
\end{proof}

Since $p<1/2$, the lower bound of Lemma \ref{lemma:01} is greater than the upper bound of Lemma \ref{lemma:02} for a sufficiently small  value of $\delta$. Thus, w.h.p.\ we have $\min_{i\in A_1} s_{1i} \ge \max_{j\not\in A_1} s_{1j}$, i.e., all the top $k$ rows chosen by $\mathtt{PAF}(k)$ are from $A_1$. In other words, if $E_{1,n}$ denotes the event that there is an error in Step 1 of the algorithm, i.e., $\mathtt{PAF}(k)$ chooses some rows from outside $A_1$, then 
\begin{align}
  \label{eq:1}
  Pr[E_{1,n}] \rightarrow 0, \text{ as } n\rightarrow \infty.
\end{align}

\begin{remark}
  The above two lemmas are valid for both case a) with $k=o(n)$ and case b) with $k=\Theta(n)$.
\end{remark}

\subsection{Analysis of Step 2 of the Algorithm}  
\label{subsec:step2}
Suppose $\alpha \in (0,1/2)$. First we condition on the event that  Step 1 does not make an error (i.e., the event $E_{1,n}^c$). 
Let $\mathcal S$ denote the set of column indices such that $\mathbf X(1,j)=1$ and $\mathbf Y(1,j)=*$, and suppose $\mathbf X_k$ and $\mathbf Y_k$ denote the sub-matrices of $\mathbf X$ and $\mathbf Y$ respectively, consisting of the  top $k$ neighbors. Also let $j_{max}$ denote the most popular column chosen by $\mathtt{PAF}(k)$, i.e., $j_{max}:=\arg\max_{j\in \mathcal S} |\mathbf Y_k(:,j)|_1$. The statistics of the columns in $\mathcal S$ are independent of the event $E_{1,n}$. Thus, conditioned on $E_{1,n}^c$, for $j\in \mathcal S$, 
 we  have $|\mathbf Y_k(:,j)|_1\sim B(k,(1-\epsilon)(1-p))$. Define $\mu_Y:=\mathbb \mathbb E[|\mathbf Y_k(:,j)|_1 ]$ and $\sigma_Y^2:=Var(|\mathbf Y_k(:,j)|_1 )$. We note that because of the i.i.d.\ nature of the columns of $\mathbf Y$, the mean $\mu_Y$ and the variance $\sigma_Y^2$ do not depend on $j$.  We have the following lemma.

\begin{lemma}[{1's form overwhelming majority in the most popular column}]
  \label{lemma:102}
 Let $j_{max}$ be the most popular column. Under both case a) and b), there exists a sequence of positive reals $\{c_n\}$, such that $c_n\rightarrow \infty$ with $n$, and w.h.p.
\[|\mathbf Y_k(:,j_{max})|_1 -|\mathbf Y_k(:,j_{max})|_0 \ge c_n.\]
\end{lemma}
   \begin{proof}
The proof is given in Appendix \ref{proof:lemma:102}. 
\end{proof}

 Now we use Lemma \ref{lemma:102} to prove that \texttt{PAF} makes vanishingly small probability of error.  Suppose  
$$\mathcal M_n:=\{\bar y\in \{0,1,*\}^k: \left(|\bar y|_1-|\bar y|_0\right) \ge c_n\},$$
where $\{c_n\}$'s are as in Lemma \ref{lemma:102}.
 We also observe that for a column $j$, 
\begin{align}
  \label{eq:103}
  \mathbf X_k(:,j) \longrightarrow \mathbf Y_k(:,j)\longrightarrow \{j_{max}=j\},
\end{align}
i.e., the random variables $\{\mathbf X_k(:,j), \mathbf Y_k(:,j), \{j_{max}=j\}\}$ form a Markov chain.
We are interested in finding the overall probability of error. 
Due to the i.i.d.\ nature of the data model, all the  columns of $\mathbf X$ have same distribution. Thus we have 
\begin{align}
  \label{eq:1001}
\nonumber  P_e[\mathtt{PAF}(k)] & = \sum_{j=1}^n Pr[j_{max}=j]\cdot Pr[\mathbf X(1,j)=0\big |j_{max}=j]\\
  & =  Pr[\mathbf X(1,1)=0\big |j_{max}=1].
\end{align}
Here on, we analyze the error probability conditioned on the event that $j_{max}=1$.
 In the following, by $p_{k,j}(\bar y)$ we mean $Pr[\mathbf Y_k(:,j)=\bar y|j_{max}=j, E_{1,n}^c]$. Thus
\begin{align}
\nonumber  P_e[\mathtt{PAF}(k)]    = & Pr[\mathbf X(1,1)=0|j_{max}=1]\\
\nonumber   \buildrel(a)\over = & Pr[\mathbf X(1,1)=0|j_{max}=1, E_{1,n}^c] +o(1)\\
\nonumber   =& \sum_{\substack{\bar y \in \{0,1, *\}^k}} Pr[\mathbf X(1,1)=0, \mathbf Y_k(:,1)=\bar y|j_{max}=1, E_{1,n}^c]+o(1)\\
\nonumber   \buildrel (b)\over = & \sum_{\bar y \in \mathcal M_n} Pr[\mathbf X(1,1)=0, \mathbf Y_k(:,1)=\bar y|j_{max}=1, E_{1,n}^c] + o(1)\\
\nonumber   \buildrel (c)\over = &\sum_{\bar y \in \mathcal M_n} Pr[\mathbf X(1,1)=0\big|\mathbf Y_k(:,1)=\bar y, E_{1,n}^c]\cdot p_{k,1}(\bar y) + o(1)\\
\nonumber   \buildrel (d)\over = &\sum_{\bar y \in \mathcal M_n} \frac{Pr[\mathbf Y_k(:,1)=\bar y\big|\mathbf X(1,1)=0, E_{1,n}^c]}{2Pr[\mathbf Y_k(:,1)=\bar y|E_{1,n}^c]} p_{k,1}(\bar y) + o(1)\\
\nonumber   =&\sum_{\bar y \in \mathcal M_n} \frac{p^{|\bar y|_1}(1-p)^{|\bar y|_0}}{p^{|\bar y|_1}(1-p)^{|\bar y|_0} + p^{|\bar y|_0}(1-p)^{|\bar y|_1}} p_{k,1}(\bar y) + o(1)\\
\nonumber  = &\sum_{\bar y \in \mathcal M_n} \frac{p^{|\bar y|_1-|\bar y|_0}}{p^{|\bar y|_1-|\bar y|_0} + (1-p)^{|\bar y|_1-|\bar y|_0}}p_{k,1}(\bar y)+ o(1)\\
\nonumber  \le &\max_{\bar y \in \mathcal M_n} \frac{p^{|\bar y|_1-|\bar y|_0}}{p^{|\bar y|_1-|\bar y|_0} + (1-p)^{|\bar y|_1-|\bar y|_0}}+ o(1)\\
\label{eq:2}  & \buildrel (e) \over = o(1),
\end{align}
where (a) follows from \eqref{eq:1}, (b) is true because  Lemma \ref{lemma:102} says that $\mathcal M_n$ happens w.h.p., (c) is due to the Markov property \eqref{eq:103} and the notation of $p_{k,j}(\bar y)$,  (d) is the Bayes' expansion, and (e) is true since for $\bar y\in \mathcal M_n$, $|\bar y|_1 -|\bar y|_0 \ge c_n$, and the fact that for $p<1/2$, $\frac{p^x}{p^x+(1-p)^x}\rightarrow 0$ ans $x\rightarrow \infty$. This proves that $P_e[\mathtt{PAF}(k)]  \rightarrow 0$. 

When $\alpha =0$ and $k$ increases to infinity with $n$, by following a similar line of statements as above, we see that there are increasingly many 1's in the most popular column, and 1's also for a majority in that column, thus the error probability approaches 0. We omit the details here.

\section{Proof of Theorem \ref{thm:02}}
\label{sec:proof:02}
The analysis for the Step 1 of the algorithm is exactly same as in Section \ref{subsec:step1}. Here we analyze the Step 2 of $\mathtt{PAF}(k)$, conditioned on the event that all the top $k$ neighbors are good (the event $E_{1,n}^c$).

Recall that $k=n^{\alpha -\gamma +g_n}$. We show that in this case the most popular column of $\mathbf Y_k$ (the top $k$ rows of $\mathbf Y$) has a finite number of  unerased entries. This allows us to find a lower bound on the probability of error. Let $\mathcal H$ denote the set of column indices such that $\mathbf Y(1,j)=*$, i.e., the columns where entries of the first row are ``hidden''.
\begin{lemma}[{Finite  number of unerased entries}]
  \label{lemma:103}
W.h.p. $$\max_{j\in \mathcal H} |\mathbf Y_k(:,j)| \le \lfloor 1/\gamma\rfloor.$$ 
\end{lemma}
\begin{proof}
The  proof is given in Appendix \ref{proof:lemma:103}. 
\end{proof}

\subsection{When $1/\gamma$  is Not an Integer}
\label{subsec:non-integer}
As in the previous section, let $j_{max}$ be the column that is recommended by the $\mathtt{PAF}$ algorithm. Due to Lemma \ref{lemma:103}, we have $|\mathbf Y_k(:,j_{max})|_1 \le \lfloor 1/\gamma \rfloor$ w.h.p.. When $1/\gamma$ is not an integer, the following lemma says that w.h.p. it is infact equal to $\lfloor 1/\gamma \rfloor$, i.e., in the most popular column, all the observed entries are 1's.

\begin{lemma}
  \label{lemma:non-integer}
  If $\lfloor 1/\gamma \rfloor$ is not an integer, then w.h.p.
$$|\mathbf Y_k(:,j_{max})|_1 = \lfloor 1/\gamma\rfloor, \text { and } |\mathbf Y_k(:,j_{max})|_0=0.$$
\end{lemma}
\begin{proof}
  The proof is given in Appendix \ref{proof:lemma:non-integer}.
\end{proof}

Suppose $$\mathcal I_n:=\{\bar y \in \{0,1,*\}^k : |\bar y|_1= \lfloor 1/\gamma\rfloor, \text{ and } |\bar y|_0=0 \}.$$
Lemma \ref{lemma:non-integer} says that $\mathcal I_n$ happens with high probability.
 We want to find the limiting behavior of the  total probability of error. By following the steps as in \eqref{eq:2} and replacing the event $M$ by the event $I$ (this replacement is justified due to Lemma \ref{lemma:non-integer}), we have 
\begin{align*}
 P_e[\mathtt{PAF}(k)] 
 = &\sum_{\bar y \in \mathcal I_n} \frac{p^{|\bar y|_1-|\bar y|_0}}{p^{|\bar y|_1-|\bar y|_0} + (1-p)^{|\bar y|_1-|\bar y|_0}}p_{k,1}(\bar y)+ o(1)\\
\buildrel (a) \over  = & \frac{p^{\lfloor \frac{1}{\gamma}\rfloor}}{p^{\lfloor \frac{1}{\gamma}\rfloor} + (1-p)^{\lfloor \frac{1}{\gamma}\rfloor}}  \sum_{\bar y \in \mathcal I_n} p_{k,1}(\bar y)+ o(1)\\
 \buildrel (b) \over =  &\frac{p^{\lfloor \frac{1}{\gamma}\rfloor}}{p^{\lfloor \frac{1}{\gamma}\rfloor} + (1-p)^{\lfloor \frac{1}{\gamma}\rfloor}} (1-o(1)) +o(1)\\
   = &\frac{p^{\lfloor \frac{1}{\gamma}\rfloor}}{p^{\lfloor \frac{1}{\gamma}\rfloor} + (1-p)^{\lfloor \frac{1}{\gamma}\rfloor}}+o(1),\\
\end{align*}
where (a) is true due to the definition of the set $I$, and (b) is true because of Lemma \ref{lemma:non-integer} ($\mathcal I_n$ happens w.h.p.). Thus we have proved that
$$\lim_{n\rightarrow \infty} P_e[\mathtt{PAF}(k)]  = \frac{p^{\lfloor \frac{1}{\gamma}\rfloor}}{p^{\lfloor \frac{1}{\gamma}\rfloor} + (1-p)^{\lfloor \frac{1}{\gamma}\rfloor}}.$$

\subsection{When $1/\gamma$ is an Integer}
\label{subsec:integer}
We first prove the lower bound on the probability of error. Due to Lemma \ref{lemma:103}, we have that $|\mathbf Y_k(:,j_{max})|_1 \le 1/\gamma $ w.h.p.. 
Define $$\mathcal J_n:=\{\bar y \in \{0,1,*\}^k : |\bar y|\le 1/\gamma \}.$$
Thus Lemma \ref{lemma:103} says that $\mathcal J_n$ happens with high probability.
By following the steps as in \eqref{eq:2} and replacing the event $M$ by the event $J$ (this replacement is justified due to Lemma \ref{lemma:103}), we have 
\begin{align*}
  P_e[\mathtt{PAF}(k)]   = &\sum_{\bar y \in \mathcal J_n} \frac{p^{|\bar y|_1-|\bar y|_0}}{p^{|\bar y|_1-|\bar y|_0} + (1-p)^{|\bar y|_1-|\bar y|_0}}p_{k,j}(\bar y)+ o(1)\\
 \ge &\min_{\bar y \in \mathcal J_n} \frac{p^{|\bar y|_1-|\bar y|_0}}{p^{|\bar y|_1-|\bar y|_0} + (1-p)^{|\bar y|_1-|\bar y|_0}} \sum_{\bar y \in \mathcal J_n} p_{k,j}(\bar y) + o(1)\\
& \buildrel  (a) \over = \min_{\bar y \in \mathcal J_n} \frac{p^{|\bar y|_1-|\bar y|_0}}{p^{|\bar y|_1-|\bar y|_0} + (1-p)^{|\bar y|_1-|\bar y|_0}} (1-o(1)) + o(1)\\
& \buildrel (b) \over\ge \frac{p^{\frac{1}{\gamma} }}{p^{ \frac{1}{\gamma} } + (1-p)^{\frac{1}{\gamma} }} (1-o(1))+o(1)\\
  &\quad  = \frac{p^{\frac{1}{\gamma} }}{p^{  \frac{1}{\gamma} } + (1-p)^{  \frac{1}{\gamma} }}+o(1),\\
\end{align*}
where (a) is true because of Lemma \ref{lemma:103}, and (b) is true  since  $|\bar y|_1 - |\bar y|_0 \le |\bar y| \le   1/\gamma $ for $\bar y \in I$, and for $x\in \mathbb R$, $\frac{p^{x}}{p^{x}+ (1-p)^{x}}$ is a decreasing function of $x$ for $p<1/2$. Thus we have
\begin{align}
  \label{eq:lower-bound}
  \lim \inf_{n\rightarrow \infty} P_e[\texttt{PAF}(T)]   \ge \frac{p^{  \frac{1}{\gamma} }}{p^{  \frac{1}{\gamma} } + (1-p)^{  \frac{1}{\gamma} }},
\end{align}
which proves the lower bound. To prove the upper bound, we need the following lemma.
\begin{lemma}
  \label{lemma:integer}
If $1/\gamma$ is an integer,  we have w.h.p. 
$$|\mathbf Y_k(:,j_{max})|_1-|\mathbf Y_k(:,j_{max})|_0\ge 1/\gamma -1.$$
\end{lemma}
\begin{proof}
  The proof is given in Appendix \ref{proof:lemma:integer}.
\end{proof}
Define $$\mathcal K_n:=\{\bar y \in \{0,1,*\}^k: |\bar y|_1 -|\bar y|_0 \ge 1/\gamma -1\}.$$
The above lemma say that $K_n$ occurs with high probability.
Then following the steps as in \eqref{eq:2}, and due to Lemma \ref{lemma:integer}, we have 
\begin{align*}
  P_e[\mathtt{PAF}(k)]   = &\sum_{\bar y \in \mathcal K_n} \frac{p^{|\bar y|_1-|\bar y|_0}}{p^{|\bar y|_1-|\bar y|_0} + (1-p)^{|\bar y|_1-|\bar y|_0}}p_{k,j}(\bar y)+ o(1)\\
 \le &\max_{\bar y \in \mathcal K_n} \frac{p^{|\bar y|_1-|\bar y|_0}}{p^{|\bar y|_1-|\bar y|_0} + (1-p)^{|\bar y|_1-|\bar y|_0}} + o(1)\\
 \buildrel (a) \over\le &\frac{p^{  \frac{1}{\gamma}-1}}{p^{  \frac{1}{\gamma} -1} + (1-p)^{  \frac{1}{\gamma} -1}} +o(1),
\end{align*}
where (a) is true because of Lemma \ref{lemma:integer}, the definition of $K$ and the observation that for $x\in \mathbb R$, $\frac{p^{x}}{p^{x}+ (1-p)^{x}}$ is a decreasing function of $x$ for $p<1/2$.  Thus we have 
$$\lim \sup_{n\rightarrow \infty} P_e[\mathtt{PAF}(k)] \le \frac{p^{  \frac{1}{\gamma}-1}}{p^{  \frac{1}{\gamma} -1} + (1-p)^{  \frac{1}{\gamma} -1}},$$
and this together with \eqref{eq:lower-bound} proves the Theorem when $1/\gamma$ is an integer.

To prove optimality of $T=k$, we consider neighborhood sizes $T_1$ and $T_2$ such that $T_1 <k <T_2$. We then consider a related estimation problem, for which the maximum a posteriori (MAP) estimator has probability of error equal to that of the $\mathtt{PAF}(k)$. We also show that the probability of error for $\mathtt{PAF}(T_1)$ and $\mathtt{PAF}(T_2)$ equals that of two sub-optimal estimators of the above mentioned related estimation problem.
Since MAP estimator minimizes probability of error over all estimators \cite[p. 8]{Poor1}, this would prove the lemma.  
The detailed proof is presented in Appendix \ref{proof:lemma:k_optimal}. 

The optimality of $T=k$ shows that $\forall T$, $\lim\sup_{n\rightarrow \infty} P_e[\mathtt{PAF}(k)] \le \lim\inf_{n\rightarrow \infty} P_e[\mathtt{PAF}(T)].$ By substituting $T=k$, we obtain
\[\lim\sup_{n\rightarrow \infty} P_e[\mathtt{PAF}(k)]\le \lim\inf_{n\rightarrow \infty} P_e[\mathtt{PAF}(k)]\le  \lim\sup_{n\rightarrow \infty} P_e[\mathtt{PAF}(k)].\]
Thus the limit $\lim_{n\rightarrow \infty}P_e[\mathtt{PAF}(k)]$ exists.
This completes the proof of Theorem \ref{thm:02}.

\section{Proof of Theorem \ref{thm:03}}
\label{sec:proof:03}
Assume that $\alpha = \frac{1}{2}+\beta$, with $\beta > 0$. We assume that there are no errors (only erasures). i.e., $p=0$, and show that the algorithm fails. To start with, we show that  w.h.p. every row overlaps with the first row at most a finite number of places. This in turn implies that amongst the top $T$ neighbors, only a vanishingly small fraction are good neighbors. Recall the definition of $s_{ij}$ from \eqref{eq:def:01} that measures the similarity between two rows.

\begin{lemma}[{Finite overlap}]
  \label{lemma:04}
There exists a constant $t_{max}>0$ (which depends on $\beta$) such that 
w.h.p. $max_{i\neq 1}  s_{1i} \le t_{max}.$
\end{lemma}
\begin{proof}
The proof is given in Appendix \ref{proof:lemma:04}.  
\end{proof}
Using Lemma \ref{lemma:04}, we first show that most neighbors of row 1 are bad.

\subsection{Most Neighbors are bad}
Suppose for a non-negative integer $m$, $N_{good}(m)$ denotes the number of neighbors (excluding row 1 itself) from $A_1$ that has $m$ commonly sampled entries with row 1, i.e., 
\begin{equation}
\label{eq:def:03}N_{good}(m):= |\{i\in A_1: s_{1i} = m\}|.
\end{equation}
More generally, for a row cluster $A_i$, we define 
\begin{equation}
\label{eq:def:04}N_i(m):=|\{j \in A_i: s_{1j} =m\}|, 
\end{equation}
to be the number of neighbors in $A_i$ with $m$ commonly sampled entries. We see that $N_1(m)=N_{good}(m)$. The total number of neighbors outside $A_1$ are denoted by 
\begin{equation}
\label{eq:def:05}N_{bad}(m):=N_2(m)+...+N_r(m).
\end{equation}
Let 
\begin{equation}
\label{eq:def:06}N(m):=N_{good}(m)+N_{bad}(m)
\end{equation}
denote the total  number of  neighbors.  We show that for all $m\le t_{max}$, $N_{good}(m)$ forms a vanishingly small fraction of $N(m)$. In the following lemma, we show that for ``large'' values of $k$, w.h.p. all the row clusters contribute equally to the top $T$ neighbors (upto a constant factor), and for ``moderate'' values of $k$, w.h.p. the contribution of the first row cluster is vanishingly small compared to the total contribution of the other row clusters, and for ``small'' values of $k$, w.h.p. the first row cluster does not contribute to the top $T$ neighbors. For all the three cases, amongst the top $T$ neighbors, w.h.p. we have vanishingly small number good neighbors compared to the bad neighbors. 
 \begin{lemma}[{Most neighbors are bad}]
   \label{lemma:const_ratio}
  There exists a constant $c_4>0$ such that for $m=1,2,...,t_{max}$,
   \begin{enumerate}
   \item If $k > c_4n^{m(2\alpha -1)}\log r$, then w.h.p. 
$$\{N_j(m)\}_{j=1}^r= \Theta\left(\mathbb E[N_1(m)]\right)=\Theta\left(\frac{k}{n^{m(2\alpha-1)}}\right).$$
   \item If there exists a constant $c_5 > 0$ such that  $c_5n^{m(2\alpha -1)} \le k \le c_4n^{m(2\alpha -1)}\log r$, then w.h.p.
     $$\{N_j(m)\}_{j=1}^r=O(\log r).$$ Moreover, there exists a subset $S$ of $[r]\backslash \{1\}$ such that $|S|=\Omega(r)$, and for all $j\in S$ we have $N_j(m)\ge 1$.
   \item If $k=o(n^{m(2\alpha -1)})$, then w.h.p. $N_{good}(m)=0$.
   \end{enumerate}
 \end{lemma}
 \begin{proof}
   The  proof is given in Appendix \ref{proof:lemma:const_ratio}.
\end{proof}

Since $N_{bad}(m)=\sum_{i=2}^r N_i(m)$ and $r$ goes to infinity with $n$, Lemma \ref{lemma:const_ratio} implies that good neighbors form a vanishingly small fraction of the total number of neighbors.
Let $N_j(m^+)$ denote the number of neighbors from the cluster $A_j$ with an overlap at  $m$ or more entries. In other words, 
\begin{equation}
\label{eq:label:07}N_j(m^+):= \sum_{t=m}^{t_{max}} N_j(t).
\end{equation}
Also let $N(m^+)$ denote the total number of neighbors with an overlap of more than or equal to $m$ entries, i.e., 
\begin{equation}
\label{eq:def:08}N(m^+):=\sum_{j=1}^r N_j(m^+). 
\end{equation}
Lemma \ref{lemma:const_ratio} implies the following corollary.

\begin{corollary}
  \label{cor:const_ratio}
  There exists a constant $c_4>0$ such that for $m=1,2,...,t_{max}$,
   \begin{enumerate}
   \item If $k > c_4n^{m(2\alpha -1)}\log r$, then w.h.p. 
$$\{N_j(m^+)\}_{j=1}^r= \Theta\left(\frac{k}{n^{m(2\alpha-1)}}\right).$$
   \item If there exists a constant $c_5>0$ such that $c_5n^{m(2\alpha -1)} \le k \le c_4n^{m(2\alpha -1)}\log r$, then w.h.p.
     $$\{N_j(m^+)\}_{j=1}^r=O(\log r).$$
Moreover, there exists a subset $S$ of $[r]\backslash \{1\}$ such that $|S|=\Omega(r)$, and for all $j\in S$ we have $N_j(m^+)\ge 1$.
   \item If $k=o(n^{m(2\alpha -1)})$, then w.h.p. $N_{good}(m^+)=0$.
   \end{enumerate}
\end{corollary}
\begin{proof}
  The proof is given in Appendix \ref{proof:cor:const_ratio}.
\end{proof}

\subsection{Even the Top Few Neighbors are Mostly bad}
Now we analyze what happens when we pick the top $T$ rows (neighbors). We show that even amongst the top $T$ neighbors, only a vanishingly small fraction are good neighbors. 

Recall that $\mathbf Y_T$ denotes the $T\times n$ sub-matrix of $\mathbf Y$ obtained by picking the top $T$ neighboring rows.  Let $T_i$ denote the number  of rows picked from the cluster $i$ (excluding the first row itself). Thus $T=\sum_{i=1}^r T_i+1$. Suppose $m_0$  is a positive integer such that
 \begin{align}
\label{eq:323}N((m_0+1)^+) < T \le  N(m_0^+).
\end{align}
Then amongst the top $T$ neighbors, we have all the rows that  overlap at $m_0+1$ positions or more, and some of the rows that overlap at  $m_0$ entries.  To be precise,
\begin{align}
  \label{eq:324}
  T_i=N_i((m_0+1)^+) + \xi_i,
\end{align}
where $\xi_i$ is a hyper-geometric random variable with parameters $(N(m_0), N_i(m_0), T-1-N((m_0+1)^+) )$  \footnote{After picking all the neighbors with an overlap of $m_0+1$ or more places, we need to pick $T-1-N((m_0+1)^+)$ more neighbors with an overlap of $m$ positions. But there are $N(m_0)$ neighbors with an overlap of $m_0$ positions, out of which $N_i(m_0)$ are from the cluster $i$.  See Appendix \ref{app:hyper} for the definition of a hyper-geometric  random variable and some useful tail bounds.}, implying 
\begin{align}
\label{eq:325}
\mathbb E[\xi_i]= \frac{N_i(m_0)}{N(m_0)}(T-1-N((m_0+1)^+)).
\end{align}
Summing both the sides of \eqref{eq:324} over $i$, we observe that 
\begin{align}
\label{eq:325:1}\sum_{i=1}^r \xi_i =T-1-N((m_0+1)^+).
\end{align}
From (\ref{eq:324}) and (\ref{eq:325}) we obtain
\begin{align}
  \label{eq:327}  \mathbb E[T_i]& = N_i((m_0+1)^+) + \frac{N_i(m_0)}{N(m_0)}(T-1-N((m_0+1)^+)).
\end{align}
Lemma \ref{lemma:const_ratio} and Corollary \ref{cor:const_ratio} now imply that that $\mathbb E[T_1]$ forms a vanishingly small fraction of of $T$. Using the Chvatal's hyper-geometric concentration lemma (see Lemma \ref{lemma:chvatal} in Appendix \ref{app:hyper}), we show in the following lemma that  this is not just true for the expectation, but w.h.p. also for  $T_1$.

\begin{lemma}[{Top neighbors are bad too}]
  \label{lemma:topT}
There is a positive integer $d$ and positive constants $c_6, c_7$, such that depending on the value of $T$, w.h.p. one of the following occurs.
\begin{enumerate}
\item[$(C_1)$] $T_1 > c_6 \log r$, and for  $i=2, 3, ... , r$ we have $dT_i \ge  T_1$.
\item[$(C_2)$] $0<T_1 \le c_6\log r$, and there is a subset $S$ of $[r]\backslash \{1\}$ with $|S|\ge c_7\frac{r}{\log^2 r}$ such that $\forall i\in S$ we have $T_i \ge 1$. 
\item[$(C_3)$]  $T_1=0$. 
\end{enumerate}
\end{lemma}

\begin{proof}
 The proof is given in Appendix \ref{proof:lemma:topT}.
\end{proof}
This implies that amongst the top $T$ neighbors, only a vanishingly small fraction are good neighbors.  Step 2 of the $\mathtt{PAF}$ algorithm now performs a majority decoding on $\mathbf Y_T$, i.e., it recommends a column  $$j_{max}=\arg\max_{j: \mathbf Y_T(1,j)=*} |\mathbf Y_T(:,j)|_1,$$ leading to a probability of error $P_e^{maj}[\mathbf Y_T]:= Pr[\mathbf X(1,j_{max})=0]$.  Thus we have
\begin{align}
  \label{eq:maj1}
   P_e\left[\mathtt{PAF}(T)\right]= P_e^{maj}\left[\mathbf \mathbf Y_T\right].
\end{align}
 In the following section, we show that probability of error for the majority decoding approaches 1/2 w.h.p..

\subsection{Analysis of Step 2 of the Algorithm}
In this section, we show that since the top $T$ rows include many bad rows, choosing the most popular item amongst the top $T$ rows does not perform well. To this end, since direct calculations are not analytically tractable, we take a somewhat circuitous route.  We first show that when we increase the number of good neighbors and decrease the number of bad neighbors in a certain way, and some of the missing entries are revealed, then the probability of error reduces. We then lower bound the probability of error for this modified case, which is easier to analyze.
We first introduce a new notation to represent the class of binary matrices with non-uniform cluster sizes.
Suppose $\mathbf a$ and $\mathbf b$ are two  vectors of length $r$. 
\begin{definition}[{Random binary matrix}]
\label{def:matrix}
  Let  $\mathbf X$ be a binary block constant matrix, whose $i$th row cluster $A_i$ is of size $\mathbf a(i)$  and the $j$th column cluster $B_j$  is of size $\mathbf b (j)$. Suppose the entries of the matrix are filled as below. If $(p,q)\in A_i\times B_j$, then $\mathbf X(p,q)=\chi_{ij}$ where $\{\chi_{ij}\}_{i,j=1}^r$ are i.i.d. Bernoulli(1/2). This class of random binary matrices is  denoted as $\mathbf X\in M_R(\mathbf a, \mathbf b)$.
\end{definition}

First we condition on the event that w.h.p. $T_1=0$ (i.e., condition $(C_3)$ of Lemma \ref{lemma:topT} is true). In this case, we see that the outcome of the majority decoding is independent of $A_1$, and hence we have
\begin{align}
\label{eq:C3} P_e^{maj}\left[\mathbf Y_T\big | C_1\right]=1/2.
\end{align}

We now consider the cases when either of the conditions $(C_1)$ or $(C_2)$ of Lemma \ref{lemma:topT} are true. For this we consider a different matrix which has more good neighbors and fewer bad neighbors compared to $\mathbf Y_T$. 
Let $u_n$ be the smallest multiple of $d$ greater than or equal to $T_1$, i.e.,
$$u_n:=d\left\lceil \frac{T_1}{d}\right\rceil,$$
and suppose there is a subset $S$ of $[r]\backslash \{1\}$ such that w.h.p.  for $j\in S$, $T_j\ge l_n$ (we have such lower bounds on $T_j$, due to Lemma \ref{lemma:topT}).
Let $\mathbf a_e$ (subscript ``e'' is for extreme values of the row cluster sizes) be the vector such that $\mathbf a_e(1)=u_n+1$,  $\mathbf a_e(j)=l_n$ for $j\in S$, and $ \mathbf a_e(j)=0$ otherwise. Also let $b_U$ be the $r$-length vector with all the  entries equal to $k$.

Suppose $\mathbf X^{(e)}\in M_R(\mathbf a_e, \mathbf b_U)$, and  {\it only the first row} of this matrix  is passed through a memoryless erasure channel with erasure probability $\epsilon$ to obtain the matrix $\mathbf Y^{(e)}$. We note that there are no erasures in the rows other than the first one. We now perform a majority decoding for $\mathbf Y^{(e)}$, and let $j_{maj}(\mathbf Y^{(e)})$ and $P_e^{maj}[\mathbf Y^{(e)}]$ be the column selected by the majority decoder, and  the corresponding probability of error respectively.
We then have the following lemma.
\begin{lemma}
  \label{lemma:maj}
For $\mathbf Y^{(e)}$ as defined above, we have
$$P_e^{maj}[\mathbf Y_T] \ge P_e^{maj}[\mathbf Y^{(e)}].$$
\end{lemma}
\begin{proof}
  The proof is given in Appendix \ref{proof:lemma:maj}.
\end{proof}

We now analyze the majority decoding on the matrix $\mathbf Y^{(e)}$, when one of the conditions $(C_1)$ or $(C_2)$ of Lemma \ref{lemma:topT} is true. We state this in the following lemma. 
\begin{lemma}
  \label{lemma:maj_unif}
  $ P_e^{maj}[\mathbf Y^{(e)} | C_1]  = 1/2 -o(1), \text{ and }  P_e^{maj}[\mathbf Y^{(e)} | C_2]= 1/2 -o(1).$
\end{lemma}
\begin{proof}
The proof is given in Appendix \ref{proof:lemma:maj_unif}. 
\end{proof}
Due to \eqref{eq:maj1} and Lemma \ref{lemma:maj}, we see that 
\begin{align}
  \label{eq:maj_final}
  \nonumber P_e[\mathtt{PAF}(T)]  &\ge P_e^{maj}[\mathbf Y^{(e)}] \\
\nonumber  & \buildrel (a)\over = \sum_{i=1}^3 P_e^{maj}[\mathbf Y^{(e)}\big|C_i] Pr[C_i]+o(1)\\ 
  & \buildrel (b) \over = 1/2+o(1).
\end{align}
where (a) is due to Lemma \ref{lemma:topT} which says that $\cup_{i\in\{1,2,3\}}C_i$ occurs w.h.p., and (b) follows from \eqref{eq:C3} and Lemma \ref{lemma:maj_unif}.
This completes the proof of  Theorem \ref{thm:03}.

\section{Proof of Theorem \ref{thm:04}}
\label{sec:proof:04}
{\bf Proof of the converse:} To prove this lower bound, we first assume that an oracle tells us the true row and column clusters (i.e., $\mathcal A$ and $\mathcal B$). Let $P_{e, oracle}(n)$ denote the error probability of the MAP estimator, when we know the clusters. Thus $P_{e,oracle}(n)$ is a lower bound on the error probability of any recommender. As before, we assume wlog that we want to recommend an item to user 1 in $A_1$. 

Since entries across clusters are i.i.d., the MAP estimator would choose an item from the column cluster $B_j$ for which we have maximum number of 1's in the cluster $A_1\times B_j$ of $\mathbf Y$. We note that while \texttt{PAF} picks a maximum weight column, this algorithm picks a maximum weight cluster and recommends a movie from that cluster. Because of the i.i.d.\ nature of the data model, the analysis for this algorithm is similar to that of analyzing \texttt{PAF}.

Suppose $\mathbf Y_{A_i\times B_j}$ denotes the matrix $\mathbf Y$ restricted to the cluster $A_i\times B_j$. By using the steps similar to those used in proving Lemma \ref{lemma:103}, we obtain that 
\[\max_{j\in\{1,2,...,r\}} |\mathbf Y_{A_1\times B_j}|\le \lfloor 1/\gamma \rfloor.\]
Then by defining 
\[\mathcal L_n:=\{\bar y\in \{0,1,*\}^{k\times k}: |\bar y| \le \lfloor 1/\gamma \rfloor\},\]
and using the steps similar to those used for proving the lower bound in  Section \ref{subsec:integer}, we see that 
\[P_{e,oracle}(n) \ge \frac{p^{\lfloor \frac{1}{\gamma}\rfloor}}{p^{\lfloor \frac{1}{\gamma}\rfloor} + (1-p)^{\lfloor \frac{1}{\gamma}\rfloor}} + o(1).\] 
Since $P_{e,oracle}(n)$ is a lower bound for the error probability of any recommender, we have 
\[P_{e}(n)\ge P_{e,oracle}(n) \ge \frac{p^{\lfloor \frac{1}{\gamma}\rfloor}}{p^{\lfloor \frac{1}{\gamma}\rfloor} + (1-p)^{\lfloor \frac{1}{\gamma}\rfloor}} + o(1).\] 
Taking $\lim\inf_{n\rightarrow \infty}$ of both the sides proves the converse.

{\bf Proof of  achievability:} We want to recommend an item to user 1 in row cluster $A_1$. We use the following algorithm to achieve the bounds. First we cluster the rows and the columns of the matrix as below. Each row chooses the $k$ most similar rows, and each column chooses the $k$ most similar columns (see Section \ref{subsec:locPop} for the definition of ``similarity''). For $\alpha <1/2$, below we show that  all the rows (or columns) find the right set of neighbors, and thus we can find the true clusters of the matrix. Let the row clusters be denoted by $A_i$'s, while $B_j$'s denote the column clusters. To recommend, we choose an (unseen) item from the column cluster $B_j$ for which we have the maximum number of 1's in the cluster $A_1\times B_j$. Let $P_e(n)$ denote the probability of error for this algorithm.

First we show that indeed w.h.p.\ the above method leads to correct clustering of the matrix. Let $E_{3,n}$ denote the event that we make an error in clustering. For a row cluster $A_i$, let $\mathbf X_{A_i}$ denote the matrix $\mathbf X$ restricted to the rows in $A_i$. For row clusters $A_i$ and $A_j$, suppose $D_{ij}$ denotes the number of column clusters at which $\mathbf X_{A_i}$ and $\mathbf X_{A_j}$ differ. Then for $i\neq j$, $D_{ij}\sim B(r,1/2)$, and the Chernoff bound  \cite[Theorem 1.1]{Dubhashi1} implies that for $\delta \in (0,1)$ we have $Pr[D_{ij}\le \frac{r}{2}(1-\delta)]\le e^{-r\delta^2/4}$. Thus using the union bound, we obtain 
\[Pr\left[\min_{i\neq j} D_{ij}\le \frac{r}{2}(1-\delta)\right]\le r^2e^{-r\delta^2/4},\]
which approaches 0 if $r\rightarrow \infty$. Thus w.h.p.\ all the $D_{ij}$'s are greater than $\frac{r}{2}(1-\delta)$. Now using arguments similar to those used in proving Lemma \ref{lemma:01} and Lemma \ref{lemma:02} along with the union bound, we observe that w.h.p.\ all the rows find the right set of neighbors. Similarly, we can also prove that w.h.p.\ all the columns find the right set of neighbors. In other words,  $Pr[E_{3,n}^c]\rightarrow 1$ as $n\rightarrow \infty$. 

For the rest of the proof, we condition on $E_{3,n}^c$. Since $E_{3,n}^c$ happens w.h.p., wlog we can assume that the statistics of the individual clusters do not change asymptotically conditioned on $E_{3,n}^c$ (i.e., they are still i.i.d. as in the original data model). This is because if $Pr[A_n]\rightarrow 1$, and $Pr[B_n]\rightarrow b$,  then $Pr[B_n|A_n]\rightarrow b$ as well.

Once we know the clusters, we recommend an item from the column cluster $B_j$ for which we have  the maximum number of 1's in the cluster $A_1\times B_j$. Suppose we denote this item by $j_0$.
As before, suppose $\mathbf Y_{A_i\times B_j}$ denotes the matrix $\mathbf Y$ restricted to the cluster $A_i\times B_j$. 

For $k^2\ge n^{\alpha - \gamma_n}$, using steps similar to those used in proving Lemma \ref{lemma:102}, we see that there exists a sequence of positive reals ${c_n}$, such that $c_n\rightarrow \infty$, and w.h.p, 
\[|\mathbf Y_{A_1\times B_{j_0}}|_1 - |\mathbf Y_{A_1\times B_{j_0}}|_0 \ge c_n.\]
In other words, the chosen cluster has overwhelming number of 1's compared to 0's.
Then using the steps very similar to those in Section \ref{subsec:step2}, we see that 
\[P_e(n)\rightarrow 0.\]

For $k^2=n^{\alpha-\gamma + g_n}$, using the steps similar to those used in proving Lemma \ref{lemma:103}, we obtain that 
\[\max_{j\in\{1,2,...,r\}} |\mathbf Y_{A_1\times B_j}|\le \lfloor 1/\gamma \rfloor.\]
Then by defining 
\[\mathcal L_n:=\{\bar y\in \{0,1,*\}^{k\times k}: |\bar y| \le \lfloor 1/\gamma \rfloor\},\]
and using the steps similar to those used for proving the  bounds in  Section \ref{subsec:non-integer} and Section \ref{subsec:integer}, we see that 
$$ \frac{p^{\left\lfloor \frac{1}{\gamma}\right\rfloor}}{p^{\left\lfloor \frac{1}{\gamma}\right\rfloor} + (1-p)^{\left\lfloor \frac{1}{\gamma}\right\rfloor}} 
\le \lim_{n\rightarrow \infty} P_{e}(n) \le 
\frac{p^{\left\lceil \frac{1}{\gamma} -1\right\rceil}}{p^{\left\lceil \frac{1}{\gamma}-1\right\rceil} + (1-p)^{\left\lceil \frac{1}{\gamma}-1 \right\rceil}}.$$
Note that the above lower bound and the upper bound match, unless $1/\gamma$ is an integer.
This proves the achievability.

\section{Conclusion}
\label{sec:con}
We have considered a neighborhood based method (the $\mathtt{PAF}$ algorithm) for recommending items to users when some ratings are available. On MovieLens data and a snapshot of Netflix data, the BER of the \texttt{PAF} algorithm is similar to that of OptSpace\cite{Montanari1}, a method based on low-rank matrix completion. To explain this performance, we analyzed the $\mathtt{PAF}$ algorithm for a binary random matrix model introduced in \cite{Aditya1}. We consider the probability that a given recommendation is incorrect, and we identify the  regimes where the $\mathtt{PAF}$ algorithm works well, as well as the regimes where it does not. In particular, the regime of $\alpha < 1/2$ and $k = n^{\alpha -\gamma + g_n}$ where $ \gamma > 0$ and $g_n \to 0$ seems to be the most suitable to describe the observed empirical results. Several extensions of this work are feasible, that can perhaps provide further insight into the performance on real data. 

Throughout this paper, we consider the case when \texttt{PAF} recommends only one item to each user. A natural generalization is to recommend multiple items (say, $q$ items), instead of just one. Then we are interested in the probability that $t$ ($t\le q$) of these recommended items are correct. Although, because of the dependencies among the recommended items, this is not a straightforward generalization of the analysis of this paper and is an open direction for future work.
One other important direction is to consider an alternative sampling mechanism that has ``power law" characteristics similar to that seen in real data. Another direction is to generalize the class of underlying matrices.

\appendix
\label{sec:proof_lemma}

\subsection{Proof of Lemma \ref{lemma:02}}
\label{proof:lemma:02}
For a row $i\not \in A_1$, suppose 
$D_i$ denotes the number of column clusters of $\mathbf X$ that have different values in the 1st and the $i$-th row. Then there are $r-D_i$ column clusters where the 1st and the $i$-th row of $\mathbf X$ match. Then $D_ik$ denotes the number of columns of $\mathbf X$ that have  different values in the 1st and the $i$-th row.  First we observe that there exists a constant $c_1\in (0,1)$, such that 
\begin{align}
\label{eq:proof:1}
\text{w.h.p. for all }i, D_ik \ge c_1 n.
\end{align} 
This is true when $r$ is bounded, because the $i$th row and the 1st row of $\mathbf X$ differ at atleast one column cluster, implying $D_i \ge 1$, and hence $D_i k \ge k =n/r$. Using $c_1:=1/r$ proves \eqref{eq:proof:1}. When $r\rightarrow \infty$ with $n$, we have $D_i = B(r, 1/2)$. Thus, the Chernoff bound \cite[Theorem 1.1]{Dubhashi1} imply that for any $\delta \in (0,1)$,  w.h.p.\ $\min_{i\not\in A_1} D_i\ge \frac{r}{2}(1-\delta)$. Thus, $\min_{i\not\in A_1} D_ik \ge \frac{n}{2}(1-\delta)$. Using $c_1:=\frac{1-\delta}{2}$ now completes the proof of \eqref{eq:proof:1}.

Suppose we condition on the event that for all $i$, $D_ik \ge c_1 n$. We call this the event $S_{1,n}$. If  two  given entries of $X$ match, then the corresponding entries of $\mathbf Y$ are not erased and match  with probability $(1-\epsilon)^2 ((1-p)^2+p^2)$. Similarly, if two entries of $\mathbf X$ differ, then the corresponding entries of $\mathbf Y$ are not erased and match  with probability $2(1-\epsilon)^2p(1-p)$. Thus we have $s_{1i}=B((r-D_i)k,  (1-\epsilon)^2 ((1-p)^2+p^2)) + B(D_ik, 2(1-\epsilon)^2p(1-p))$. In other words, $s_{1i}$ is a sum of $n$ independent Bernoulli trials with 
\begin{align*}
  \mathbb E[s_{1i}]& = (r-D_i)k (1-\epsilon)^2 ((1-p)^2+p^2)) + 2 D_ik (1-\epsilon)^2p(1-p))\\
  & = n^{1-2\alpha}((1-p)^2 + p^2) - D_ik (1-\epsilon)^2 ((1-p)^2 + p^2 -2p(1-p))\\
& = n^{1-2\alpha}((1-p)^2 + p^2) - D_ik n^{-2\alpha} (1-2p)^2\\
& \buildrel (a) \over \le n^{1-2\alpha}((1-p)^2 + p^2) - c_1 n^{1-2\alpha} (1-2p)^2,
\end{align*}
where (a) is true because $D_ik\ge c_1n$. Thus, due to the Chernoff bound \cite[Theorem 1.1]{Dubhashi1}, conditioned on $S_{1,n}$, we have  that w.h.p.\ $s_{1i} \le n^{1-2\alpha}((1-p)^2 + p^2) - c_1 n^{1-2\alpha} (1-2p)^2(1+\delta)$. The lemma is now proven by observing from \eqref{eq:proof:1} that $S_{1,n}$ happens w.h.p..

\subsection{Proof of Lemma \ref{lemma:102}}
\label{proof:lemma:102}
We prove the lemma by first obtaining the following two lemmas proving a  lower bound for $|\mathbf Y_k(:,j_{max})|_1$, and  an upper bound for $|\mathbf Y_k(:,j_{max})|_0$ respectively, which we prove towards the end of this section.

\begin{lemma}[{Many 1's in the most popular column}]
 \label{lemma:101}
 For different values of $k$, we have the following lower bounds on $|\mathbf Y_k(:,j_{max})|_1$.
 \begin{enumerate}
 \item If $k=n^{\alpha-\gamma_n}$ such that $\gamma_n\ge 0$ and $\gamma_n\rightarrow 0$, then w.h.p. 
$$|\mathbf Y_k(:,j_{max})|_1\ge \min\left\{\sqrt{\log n}, \frac{1}{2\gamma_n}\right\}.$$
 \item If $k=n^\alpha g_n$ for $g_n \ge 1$, then w.h.p. 
$$|\mathbf Y_k(:,j_{max})|_1\ge \max\left\{\mu_Y+\min\left\{\sigma_Y^{1/4},\sqrt{\log n}\right\} \sigma_Y,\sqrt{\log n}\right\}.$$
 \end{enumerate}
\end{lemma}

\begin{lemma}[{Few 0's in the most popular column}]
 \label{lemma:101_upper_bound}
 For different values of $k$, we have the following upper bounds on $|\mathbf Y_k(:,j_{max})|_1$.
 \begin{enumerate}
 \item If $k=n^{\alpha-\gamma_n}$ such that $\gamma_n\ge 0$ and $\gamma_n\rightarrow 0$, then w.h.p. 
$$|\mathbf Y_k(:,j_{max})|_0\le \min\left\{\frac{\sqrt{\log n}}{2}, \frac{1}{4\gamma_n}\right\}.$$
 \item If $k=n^\alpha g_n$ for $g_n \ge 1$, then w.h.p. 
$$|\mathbf Y_k(:,j_{max})|_0\le \max\left\{\mu_Y+\frac{1}{2}\min\left\{\sigma_Y^{1/4},\sqrt{\log n}\right\} \sigma_Y,\frac{\sqrt{\log n}}{2}\right\}.$$
 \end{enumerate}
\end{lemma}

These two lemmas together imply that there exist a sequence of positive reals $\{c_n\}$ such that $c_n\rightarrow \infty$ with $n$, and w.h.p.
\[|\mathbf Y_k(:,j_{max})|_1- |\mathbf Y_k(:,j_{max})|_0 \ge c_n.\] 
This proves Lemma \ref{lemma:102}. Below we prove Lemma \ref{lemma:101} and Lemma \ref{lemma:101_upper_bound}.

\noindent
{\bf Proof of Lemma \ref{lemma:101}:}
Conditioned on the event that the top $k$ neighbors picked by \texttt{PAF} are all good, $\mathbf Y_k(:,j)$ is binomially distributed for $j\in \mathcal S$. We prove the lemma by carefully lower bounding the upper tail of this binomial using a theorem on moderate deviations.
\begin{enumerate}
\item Recall that we have conditioned on the event that  all the rows in the top $k$ neighbors chosen by \texttt{PAF} are good. Suppose $k=n^{\alpha-\gamma_n}$. Recall that $\mathcal S$ denotes the set of column indices such that $\mathbf X(1,j)=1$ and $\mathbf Y(1,j)=*$. 

\begin{claim}
\label{claim:01}
There exist a constant $c_2>0$, such that w.h.p.\ $|\mathcal S|\ge c_2n$.
\end{claim}

\begin{proof}[Proof of Claim \ref{claim:01}]
  We see that  $|\mathcal S| \sim  B(M,\epsilon)$ where $M \sim k \cdot B\left(r,\frac{1}{2}\right)$.  Here $M$ denotes the number of columns of $\mathbf X$ with 1's as the true ratings of user 1. For case a), where  $r$ increase to $\infty$ with $n$ (since $k=o(n)$), due to Chernoff bound \cite[Theorem 1.1]{Dubhashi1} we have w.h.p.  $|\mathcal S|\ge n/3$. For case b), where  $k=\Theta(n)$, $r$ stays bounded (suppose $r\le r_0$ always) and since the first row of $\mathbf X$ is not all zero, we have $M\ge k\ge n/r_0$. Thus due to the Chernoff bound, we have w.hp. $|\mathcal S|\ge n/2r_0$. This proves the claim.
\end{proof}

 For a column $j\in \mathcal S$ we see that $|\mathbf Y_k(:,j)|_1\sim B(k,(1-\epsilon)(1-p))$, and they are independent for different values of $j$. Thus, for $j\in \mathcal S$,
    \begin{align}
\nonumber    Pr\big[|\mathbf Y_k(:,j)|_1\ge t\big] \ge & Pr\big[|\mathbf Y_k(:,j)|_1=t\big]\\
\nonumber \buildrel (a)\over \ge &{k \choose t} ((1-\epsilon)(1-p))^t \epsilon^{k-t}\\
\label{eq:222}       \buildrel(b)\over  \ge &\left(\frac{k}{t}\right)^t \left(\frac{c(1-p)}{n^\alpha}\right)^t e^{-2\ln (2) c/n^{\gamma_n}}, \text{ for large $n$}\\
\nonumber      \buildrel (c) \over \ge &\left(\frac{c(1-p)}{tn^{\gamma_n}}\right)^t e^{-2\ln (2) c}.
    \end{align}
    where (a) is true since $1-(1-\epsilon)(1-p) \ge \epsilon$, (b) follows since $\epsilon=1-c/n^\alpha$, $1-x \ge e^{-2\ln (2) x}$ for $x\in [0,1/2]$, and ${k \choose t} \ge \left(\frac{k}{t}\right)^t$ (see \cite[p. 434]{motwani95}), and (c) is true because $\gamma_n \ge 0$.
    Since w.h.p. $|\mathcal S|\ge c_2 n$,  we now have 
    \begin{align}
      \nonumber     Pr[|\mathbf Y_k(:,j_{max})|_1 < t]  \le & Pr\left[\max_{j\in \mathcal S} |\mathbf Y_k(:,j)|_1 < t \big| |\mathcal S|\ge c_2 n \right]+o(1)\\
      \nonumber \le & \left( 1- \left(\frac{c(1-p)}{tn^{\gamma_n}}\right)^t e^{-2\ln (2) c}\right)^{c_2n}+o(1)\\
      \label{eq:201} \le & e^{-c_2n\left(\frac{c(1-p)}{tn^{\gamma_n}}\right)^t e^{-2\ln (2) c} }+o(1)
    \end{align}
Suppose we put $t= t_0:=\min\{\sqrt{\log n}, \frac{1}{2\gamma_n}\}$. Then
$$\left(\frac{tn^{\gamma_n}}{c(1-p)}\right)^t = \frac{ n^{\gamma_n t} t^t}{(c(1-p))^t} \buildrel (a) \over \le \sqrt n  \left(\frac{\sqrt{\log n}}{c(1-p)}\right)^{\sqrt{\log n}}\hspace{-0.1in}=o(n),$$
where (a) follows since $\gamma_nt \le 1/2$ and $t\le \sqrt{\log n}$. Thus, from \eqref{eq:201} we obtain
\begin{align*}
 & Pr[|\mathbf Y_k(:,j_{max})|_1 < t_0] 
\le e^{-\frac{1}{o(1)}}+o(1)=o(1).
\end{align*}
This proves the first part of the lemma.

\item Recall that we have assumed $k=n^\alpha g_n$ for $g_n \ge 1$. By following a very similar analysis as in the first part, we see that w.h.p. $|\mathbf Y_k(:,j_{max})|_1 \ge \sqrt{\log n}$. In particular for $g_n=1$ (or equivalently for $k=n^\alpha$),  \eqref{eq:222} becomes 
\begin{align}
\nonumber   Pr\big[|\mathbf Y_k(:,j)|_1\ge t\big] 
 \ge & \left(\frac{k}{t}\right)^t \left(\frac{c(1-p)}{n^\alpha}\right)^t e^{-2\ln (2) c}\\
\label{eq:555}  & =\left(\frac{c(1-p)}{t}\right)^t e^{-2\ln (2) c}.
    \end{align}
Observe that  for two random variables $X$ and $Y$ such that $X\sim B(n_1,p)$ and $Y\sim B(n_2,p)$ with $n_1 \ge n_2$, we have $Pr[X\ge t] \ge Pr[Y\ge t]$. Thus, using \eqref{eq:555} we have 
\begin{align*}
Pr\left[|\mathbf Y_k(:,j)|_1\ge t\big|g_n\ge 1\right]  &\ge Pr\left[|\mathbf Y_k(:,j)|_1\ge t\big|g_n=1\right] \\
&\ge \left(\frac{c(1-p)}{t}\right)^t e^{-2\ln (2) c}.
\end{align*}
Hence for $t=\sqrt{\log n}$, \eqref{eq:201} has the following counterpart,
\begin{align*}
          Pr[|\mathbf Y_k(:,j_{max})|_1 < t] 
\le &  e^{-c_2n\left(\frac{c(1-p)}{t}\right)^t e^{-2\ln (2) c} }+o(1)\\
&=e^{-\frac{1}{o(1)}}+o(1)=o(1).
    \end{align*}

But in Lemma \ref{lemma:102} we need better bounds for $g_n \rightarrow \infty$, and we consider this case now. Recall that for $j\in \mathcal S$, $\mu_Y=\mathbb \mathbb E[|\mathbf Y_k(:,j)|_1 ]=c(1-p)g_n$ and $\sigma_Y^2=Var(|\mathbf Y_k(:,j)|_1 )=c(1-p)g_n(1-(1-\epsilon)(1-p))$. We define $t_n:=\min\{\sigma_Y^{1/4}, \sqrt{\log n}\}$. Then $t_n^6=o(\sigma_Y^2)$ and Theorem \ref{thm:large} implies that for a column $j\in\mathcal  S$,
\begin{align*}
  Pr[|\mathbf Y_k(:,j)|_1 > \mu_Y +t_n \sigma_Y] \doteq& Q(t_n)
  \buildrel (a)\over \doteq \frac{1}{\sqrt{2\pi}t_n}e^{-t_n^2/2}\\ 
   \ge &\frac{1}{2}\frac{1}{\sqrt{2\pi}t_n}e^{-t_n^2/2}, \text{ for large $n$}\\
  &\buildrel (b)\over = \Omega\left(\frac{1}{\sqrt{n \log n}}\right).
\end{align*}
where (a) is true because $Q(t)\doteq \frac{1}{\sqrt{2\pi}t}e^{-t^2/2}$ \cite[Lemma 1.2]{Feller1}, and (b) is true since $t_n \le \sqrt{\log n}$. Since w.h.p. $|\mathcal S|\ge c_2n$, we have
\begin{align}
\nonumber Pr[|\mathbf Y_k(:,j_{max})|_1  \le \mu_Y +t_n \sigma_Y]
\le & Pr\left[\max_{j\in\mathcal S} |\mathbf Y_k(:,j)|_1 \le \mu_Y +t_n \sigma_Y \big| |\mathcal S|\ge c_2n\right]+o(1)\\
\nonumber  \le &\left(1-\Omega\left(\frac{1}{\sqrt{n \log n}}\right)\right)^{c_2n}+o(1)\\
\nonumber \le &e^{-\Omega\left(\frac{n}{\sqrt{n\log n}}\right)}+o(1)
= o(1).
\end{align}
Thus, w.h.p. $|\mathbf Y_k(:,j_{max})|_1  \ge \mu_Y +t_n \sigma_Y$, if $g_n \rightarrow \infty$. We have already observed that w.h.p. $|\mathbf Y_k(:,j_{max})|_1 \ge \sqrt{\log n}$. Thus the lemma is implied.
\end{enumerate}

\noindent
{\bf Proof of Lemma \ref{lemma:101_upper_bound}:}
 First we condition on the event that $\mathbf X(1,j_{max})=1$. We observe that
$$|\mathbf Y_k(:,j)|_0 \longrightarrow |\mathbf Y_k(:,j)|_1 \longrightarrow \{j_{max}=j\}.$$
Then conditioned on the value of $|\mathbf Y_k(:,j_{max})|_1=t$, the distribution of $|{\mathbf Y_k(:,j_{max})}|_0$ does not depend on the fact that $j_{max}$ is the most popular column chosen by the algorithm, and hence $|{\mathbf Y_k(:,j_{max})}|_0 \sim B\left(k-t, p_0\right)$, where $p_0:=\frac{p(1-\epsilon)}{p(1-\epsilon)+\epsilon}$.  This is because for a given column $j$ of $\mathbf Y_k$, upon observing that there are exactly $t$ 1's, the other $k-t$ entries are i.i.d. with probability of 0 being $p_0$.

\begin{enumerate}
\item Suppose $k=n^{\alpha - \gamma_n}$ such that $\gamma_n\rightarrow 0$. We define $b(k,p,i):={ k\choose i} p^i (1-p)^{n-i}$ to be the $i$th binomial term, and observe that $b(k,p,i)\le \left(kpe/i\right)^i$, since ${k\choose i}\le (ke/i)^i$ (see \cite[p. 434]{motwani95}). We see that 
\begin{align*}
 Pr\left[|{\mathbf Y_k(:,j_{max})}|_0 \ge \frac{\sqrt{\log n}}{2}\right]
 = &\sum_{i=\frac{\sqrt{\log n}}{2}}^{k-t} b(k-t, p_0,i)\\
 =& \sum_{i=\frac{\sqrt{\log n}}{2}}^{2\log n}b(k-t, p_0,i) + \sum_{i=2\log n+1}^{k-t} b(k-t, p_0,i)\\
\buildrel (a)\over \le & 2\log n \cdot b\left(k-t, p_0, \frac{\sqrt{\log n}}{2}\right) + k\cdot b(k-t, p_0, 2\log n+1)\\
\buildrel (b) \over \le &2\log n \left(\frac{(k-t)p_0e}{\sqrt{\log n}/2}\right)^{\frac{\sqrt{\log n}}{2}} \hspace{-0.1in}+\hspace{-0.05in} (k-t) \left(\frac{(k-t)p_0e}{2\log n+1}\right)^{2\log n+1}\\ 
\buildrel (c)\over \le &2 \log n \left(\frac{2c'}{n^{\gamma_n}\sqrt{\log n}}\right)^{\frac{\sqrt{\log n}}{2}} \hspace{-0.1in}  + k \left(\frac{c'}{n^{\gamma_n} (2\log n+1)}\right)^{2\log n+1}\\
& = o(1).
\end{align*}
where (a) is true since $b(k,p,i)$ is a decreasing function of $i$ for $i$ more than $kp$ and we have $(k-t)p_0=o(1)$, (b) is due to the fact that $b(k,p,i)\le (kpe/i)^i$ , and (c) follows by observing that $kp_0e\le \left(\frac{c'}{n^{\gamma_n}}\right)$ for a constant $c'>0$. Thus 
w.h.p. we have $|{\mathbf Y_k(:,j_{max})}|_0  < \frac{\sqrt{\log n}}{2}$.

Now suppose $\gamma_n > \frac{1}{2\sqrt{\log n}}$. Then we see that 
\begin{align*}
  Pr\left[|{\mathbf Y_k(:,j_{max})}|_0  \ge \frac{1}{4\gamma_n}\right]  &= \sum_{i=\frac{1}{4\gamma_n}}^{k-t} b(k-t, p_0,i) \\
&   \le \sum_{i=\frac{1}{4\gamma_n}}^\infty b(k-t, p_0,i) \\
 &  \buildrel (a)\over \le \sum_{i=\frac{1}{4\gamma_n}}^\infty ((k-t) p_0e/i)^i\\  
 & \buildrel (b)\over \le  \sum_{i=\frac{1}{4\gamma_n}}^\infty \left( \frac{4c' \gamma_n}{n^{\gamma_n}}\right)^i \\
& \quad \buildrel (c)\over = \Theta\left(\left( \frac{4c' \gamma_n}{n^{\gamma_n}}\right)^{1/4\gamma_n}\right)\\
& \quad \buildrel (d) \over <  \Theta\left(\frac{(4c' \gamma_n)^{\sqrt{\log n}/2}}{n^{1/4}}\right)\\
&\quad \quad=  o(1),
\end{align*}
where (a) is true since $b(k,p,i)\le (kpe/i)^i$, (b) follows because $kp_0e\le \left(\frac{c'}{n^{\gamma_n}}\right)$ for a constant $c'$, (c) is true by observing that for $x=o(1)$, we have $\sum_{i=m}^\infty x^i = \Theta(x^m)$, and (d) follows since $\frac{1}{4\gamma_n} < \frac{\sqrt{\log n}}{2}$ whenever $\gamma_n > \frac{1}{2\sqrt{\log n}}$. 

Thus we have proved that w.h.p. 
$|{\mathbf Y_k(:,j_{max})}|_0  < \min \{ \frac{\sqrt{\log n}}{2}, \frac{1}{4\gamma_n}\}.$

\item Now we consider the other case of $k=n^\alpha g_n$ for $g_n\ge 1$. If $g_n$ is upper bounded by a constant, then arguments very similar to those used in the first part tell us that w.h.p. $|{\mathbf Y_k(:,j_{max})}|_0 <\sqrt{\log n}/2$.  

In the remaining part of the proof, we assume that $g_n\rightarrow \infty$. Recall that for a column $j$ such that $\mathbf X(1,j)=1$, we have  $\mu_Y:=\mathbb \mathbb E[|\mathbf Y_k(:,j)|_1 ]=k(1-\epsilon)(1-p)$ and $\sigma_Y^2:=Var(|\mathbf Y_k(:,j)|_1 )=k(1-p)(1-\epsilon)(1-(1-\epsilon)(1-p))$. Conditioned on the value of  $|\mathbf Y_k(:,j_{max})|_1=t$, suppose $\mu_{\bar Y}$ and $\sigma_{\bar Y}^2$ denote the conditional mean and variance of $|\mathbf Y_k(:,j_{max})|_0$.  We observe that for $t\ge  \mu_Y$ and large enough $n$,
$$\mu_{\bar Y}=(k-t)p_0\le \mu_Y,\text{ and } \sigma_{\bar Y}^2 =(k-t)p_0(1-p_0)\le 2 \sigma_Y^2.$$
Suppose $t_n:=\min\{\sigma_Y^{1/4},\sqrt{\log n}\} $. Then we have $t_n^6=o(\sigma_Y^2)$, and since w.h.p. $y_1:=|\mathbf Y_k(:,j_{max})|_1\ge \mu_Y$ (see Lemma \ref{lemma:101}), using Theorem \ref{thm:large} we obtain
\begin{align*}
&  Pr\left[|{\mathbf Y_k(:,j_{max})}|_0  > \mu_{Y}+ \frac{t_n}{2} \sigma_{Y}\right] \\
 \le &Pr\left[|{\mathbf Y_k(:,j_{max})}|_0  > \mu_{\bar Y}+ \frac{1}{2\sqrt 2} t_n \sigma_{\bar Y}\big| y_1\ge \mu_{Y}\right] +o(1)\\
&  \doteq Q\left(\frac{t_n}{2\sqrt 2 }\right)\\
&= o(1). 
\end{align*}
Thus w.h.p.  $|{\mathbf Y_k(:,j_{max})}|_0 \le \max\{\sqrt{\log n}/2, \mu_Y+\frac{t_n}{2}\sigma_Y\}$.

\begin{remark}
  In the above proof, we had conditioned on the event that $\mathbf X(1, j_{max})=1$. When we condition on  $\mathbf X(1,j_{max})=0$, we have $p_0=\frac{(1-p)(1-\epsilon)}{(1-p)(1-\epsilon)+\epsilon}$, and a very similar set of steps  prove the lemma.
\end{remark}
\end{enumerate}

\subsection{Proof of Lemma \ref{lemma:103}}
\label{proof:lemma:103}
We condition on the event $E_{1,n}^c$ that all the top $k$ rows are good . Due to \ref{eq:1}, this event $E_{1,n}^c$ occurs w.h.p.. Then we observe that for a column $j\in \mathcal H$, $|\mathbf Y_k(:,j)|\sim B(k,1-\epsilon)$. Thus 
\begin{align*}
    Pr[|\mathbf Y_k(:,j)|\ge m] &=\sum_{t=m}^k{k \choose t}(1-\epsilon)^t \epsilon^{k-t}\\
    & \le \sum_{t=m}^k \left(\frac{kc}{n^\alpha}\right)^t, \text{ since ${k\choose t} \le k^t$}\\
    & \le \sum_{t=m}^k \left(\frac{c}{n^{\gamma}}\right)^t, \text{ since $k\le n^{\alpha-{\gamma}}$}\\
        & \le \sum_{t=m}^\infty \left(\frac{c}{n^{\gamma}}\right)^t\\
        & \quad =\frac{\frac{c^m}{n^{{\gamma} m}}}{1-\frac{c}{n^{\gamma}}}
         \le 2 \frac{c^m}{n^{{\gamma} m}}, \text{ for large $n$}.
  \end{align*}
Thus we have using union bound, 
\begin{align*}
  Pr[\max_{j\in \mathcal H} |\mathbf Y_k(:,j)|  \ge m]  &\le n\cdot 2 \frac{c^m}{n^{{\gamma} m}}=2c^mn^{1-{\gamma} m}\rightarrow 0,
\end{align*}
if $m  > 1/{\gamma}$. In other words, w.h.p. we have $\max_{j\in \mathcal H} |\mathbf Y_k(:,j)| \le \lfloor 1/{\gamma}\rfloor$, conditioned on $E_{1,n}^c$. Since $Pr[E_{1,n}^c]=o(1)$, we have w.h.p.  $\max_{j\in \mathcal H} |\mathbf Y_k(:,j)| \le \lfloor 1/{\gamma}\rfloor$.

\subsection{Proof of Lemma \ref{lemma:non-integer}}
\label{proof:lemma:non-integer}
Conditioned on the event $E_{1,n}^c$, for a column $j\in \mathcal H$, $|\mathbf Y_k(:,j)|\sim B(k,1-\epsilon)$. 
Thus 
\begin{align}
  \label{eq:1002}
\nonumber  Pr\left[|\mathbf Y_k(:,j)|=\lfloor 1/\gamma\rfloor \right] & = {k\choose \lfloor 1/\gamma\rfloor} (1-\epsilon)^{\lfloor 1/\gamma\rfloor} \epsilon^{k- \lfloor 1/\gamma\rfloor}\\
\nonumber &\buildrel (a)\over = \Theta \left(\left(\frac{k}{n^\alpha}\right)^{\lfloor 1/\gamma \rfloor}\right)\\
&\buildrel (b)\over =  \Theta \left(n^{(-\gamma +g_n){\lfloor 1/\gamma \rfloor}}\right)
\end{align}
where (a) is true since for a constant $m$, ${k \choose m}=\Theta(k^m)$, $1-\epsilon = c/n^\alpha$, and $\epsilon^{k-\lfloor 1/\gamma\rfloor} \rightarrow 1$, and (b) follows since $k=n^{\alpha -\gamma +g_n}$. Let $A$ be the set of columns $j\in \mathcal H$ for which $\mathbf X(1,j)=1$ and  $|\mathbf Y_k(:,j)|=\lfloor 1/\gamma\rfloor$. For every column $j$, let 
$$\chi_j = \left\{\begin{array}{ll}
    1, &  \text{ if } j\in A\\
    0, & \text{ otherwise.}
  \end{array}\right.$$
Then by linearity of expectation, we have 
\begin{align}
  \label{eq:1003}
  \mathbb E[|A|] = \mathbb E\left[\sum_{j=1}^n \mathbb\chi_j\right]= \sum_{j=1}^n \mathbb E[\chi_j] = \sum_{j=1}^n Pr[j\in A] \buildrel (a) \over =\Theta\left(n^{1+(-\gamma +g_n){\lfloor 1/\gamma \rfloor}}\right),
\end{align}
where (a) is true due to \eqref{eq:1002}. We see that the rightmost expression in \eqref{eq:1003} increases to infinity, since $g_n=o(1)$ and $1/\gamma$ is not an integer. Moreover, for $j\in \mathcal H$, $\chi_j$'s  are independent.
Thus using the Chernoff bound we have w.h.p. 
\begin{align}
  \label{eq:1004}
  |A| =  \Theta\left(n^{1+(-\gamma +g_n){\lfloor 1/\gamma \rfloor}}\right).
\end{align}
For a column $j\in A$,
$$Pr[|\mathbf Y_k(:,j)|_1={\lfloor 1/\gamma \rfloor}]=(1-p)^{\lfloor 1/\gamma \rfloor}.$$ 
Thus there exists a column $j\in A$ with $|\mathbf Y_k(:,j)|_1={\lfloor 1/\gamma \rfloor}$ (and hence $|\mathbf Y_k(:,j)|_0=0$), with a probability not less than $\left(1-\left(1-(1-p)^{\lfloor 1/\gamma \rfloor}\right)^{|A|}\right)\rightarrow 1$. Thus we have w.h.p. $|\mathbf Y_k(:,j_{max})|_1 \ge {\lfloor 1/\gamma \rfloor}$.
But due to Lemma \ref{lemma:103}, we have w.h.p. $|\mathbf Y_k(:,j_{max})|\le {\lfloor 1/\gamma \rfloor}$. Thus we have w.h.p. 
$$|\mathbf Y_k(:,j_{max})|_1= {\lfloor 1/\gamma \rfloor}, \text{ and } |\mathbf Y_k(:,j_{max})|_0=0.$$

\subsection{Proof of Lemma \ref{lemma:integer}}
\label{proof:lemma:integer}
Conditioned on the event $E_{1,n}^c$, for a column $j\in \mathcal H$, $|\mathbf Y_k(:,j)|\sim B(k,1-\epsilon)$. 
Suppose $A$ be the set of columns $j\in \mathcal H$ for which $|\mathbf Y_k(:,j)|=1/\gamma$. Then using similar steps as in the proof of Lemma \ref{lemma:non-integer}, we obtain for a column $j$
\begin{align}
  \label{eq:1006}
  Pr[j\in A] & = \Theta \left(n^{(-\gamma +g_n){ 1/\gamma }}\right)= \Theta\left( n^{-1+g_n/\gamma}\right),
\end{align}
and by linearity of expectation, we have 
\begin{align}
  \label{eq:1007}
  \mathbb E[|A|] = \Theta \left( n^{g_n/\gamma}\right).
\end{align}
Using Lemma \ref{lemma:low_mean} with $t=n^{g_n/\gamma}\log n$, we have w.h.p. 
\begin{align}
  \label{eq:1008}
  |A| = O \left( n^{g_n/\gamma}\log n\right).
\end{align}

Now suppose $B$ denotes the set of columns $j\in \mathcal H$ for which $\mathbf X(1,j)=1$, and $|\mathbf Y_k(:,j)|=1/\gamma -1$. Then by using similar steps as above, we obtain
\begin{align}
  \label{eq:1009}
  Pr[j\in B]& =\Theta \left( n^{-1 +\gamma + g_n(1/\gamma -1)}\right),
\end{align}
and by linearity of expectation,
\begin{align}
  \label{eq:1010}
  \mathbb E[|B|]=\Theta \left( n^{\gamma + g_n(1/\gamma -1)}\right).
\end{align}
Thus using the Chernoff bound, we obtain w.h.p. 
\begin{align}
  \label{eq:1011}
  |B|=\Theta \left( n^{\gamma + g_n(1/\gamma -1)}\right).
\end{align}
For a column $j\in B$, 
\begin{align}
  \label{eq:1012}
  Pr[|\mathbf Y(:,j)|_1=1/\gamma -1]= (1-p)^{1/\gamma -1}.
\end{align}
Thus by defining 
$$C:=\{j : |\mathbf Y(:,j)|_1=1/\gamma -1, |\mathbf Y(:,j)|_0=0\},$$
 we see that for a column $j\in B$, 
$$Pr[j\in C]=(1-p)^{1/\gamma -1},$$ and by using linearity of expectation and \eqref{eq:1011},
\begin{align}
  \label{eq:1013}
\nonumber   \mathbb E[|C|] & \ge \mathbb E[|C\cap B|]\\
\nonumber & \quad = |B| (1-p)^{1/\gamma -1}\\
  &\quad =\Omega \left( n^{\gamma + g_n(1/\gamma -1)}\right),
\end{align}
which together with the Chernoff bound implies that w.h.p.
\begin{align}
  \label{eq:1014}
|C|=  \Omega \left( n^{\gamma + g_n(1/\gamma -1)}\right).
\end{align}

Thus, for the recommended column $j_{max}$, we  have the following two possibilities.
\begin{enumerate}
\item We have  $|\mathbf Y(:,j_{max})|_1=1/\gamma$. Since  w.h.p. $|\mathbf Y(:,j_{max})|\le 1/\gamma$ due to Lemma \ref{lemma:103}, we have w.h.p. 
$$|\mathbf Y(:,j_{max})|_1 - |\mathbf Y(:,j_{max})|_0=1/\gamma.$$
\item We have $|\mathbf Y(:,j_{max})|_1=1/\gamma-1$. Then either $j_{max}\in A$, or $j_{max}\in C$. From \eqref{eq:1008} and \eqref{eq:1014}, since $g_n=o(1)$ and $\gamma >0$, we see that w.h.p. $|A|$ is vanishingly small compared to $|C|$ . Thus w.h.p. $j_{max}\in C$. Thus, from the definition of $C$, we obtain
$$|\mathbf Y(:,j_{max})|_1=1/\gamma -1, \text{ and }|\mathbf Y(:,j_{max})|_0=0.$$
\end{enumerate}
These two observations together proves the lemma.

\subsection{Proof of optimality of $T=k$}
\label{proof:lemma:k_optimal}

Recall \eqref{eq:1}, which says that all the top $k$ neighbors picked by the $\mathtt{PAF}$ algorithm are good  w.h.p.. As before, let $E_{1,n}$ denote the event that a bad neighbor is picked amongst the top $k$ neighbors. For the remainder of this proof, we condition on the event $E_{1,n}^c$, i.e., all the top $k$ neighbors are good.

Throughout this proof, a column $j$ is good if  $\mathbf X(1,j)=1$, and it is  bad if $\mathbf X(1,j)=0$. Suppose $T_1 < k < T_2$, and $A^{(1)}$ denotes a set of  $T_1$ good neighbors, $A^{(2)}$ denotes the rest of the $k-T_1$ rows of $A_1$, and $B_1$ is all the rows not in $A_1$ that are picked amongst the top $T_2$ rows. We see that $|B_1|=T_2-k$. Recall that  $\mathcal H$ denotes the set of columns $j$ such that $\mathbf Y(1,j)=*$. Now suppose, we do not get to observe $\mathbf Y$; instead we get to observe the following random variables related to $\mathbf Y$.
\begin{itemize}
\item For all the columns $j\in \mathcal H$, we observe the corresponding number of 1's restricted to $A^{(1)}$ and $A^{(2)}$. To be more precise, let $\mathbf y_j^{(i)}$ denotes the $j$-th column of $\mathbf Y_{T_2}$, restricted to $A^{(i)}$. Then we observe $(|\mathbf y_j^{(1)}|_1, |\mathbf y_j^{(2)}|_1)=(s_j^{(1)}, s_j^{(2)})$ for all columns $j\in \mathcal H$. Let $\mathcal I_1$ denote this collection of observed random variables.
\item  For all the columns $j\in \mathcal H$, we also observe the corresponding number of 1's restricted to $B_1$. To be more precise, let $\mathbf y_j^{(b)}$ denotes the $j$-th column of $\mathbf Y_{T_2}$, restricted to $B_1$ (the superscript {\it b} is for bad). Then we observe $|\mathbf y_j^{(b)}|_1=s_j^{(b)}$ for all columns $j\in \mathcal H$. Let $\mathcal I_2$ denote this collection of observed random variables.
\end{itemize}
Upon observing $\mathcal I_1$ and $\mathcal I_2$, we want to find a column $j\in \mathcal H$ such that $\mathbf X(1,j)=1$. First we consider the MAP estimator for this problem, which selects  a column $j_{MAP}$ satisfying
\begin{align}
  \label{eq:890}
  j_{MAP} & := \arg \max_{j\in \mathcal H} Pr[\mathbf X(1,j)=1\big| \mathcal I_1, \mathcal I_2].
\end{align}
We again note that we get to observe only $\mathcal I_1$ and $\mathcal I_2$, not $\mathbf Y$. This MAP decoder makes an error with probability $P_e^{MAP}:=Pr[X(1,j_{MAP})\neq 1]$. We would now show that this probability of error is same as the error probability of the $\mathtt{PAF}$ algorithm with $T=k$. Amongst the columns $j\in \mathcal H$, let $\mathcal G$ denote the set good columns (with $\mathbf X(1,j)=1$) and $\mathcal B$ denote the set of bad columns (with $\mathbf X(1,j)=0$).
With this notation, conditioned on $|\mathcal G|=m$, we now have,
\begin{align}
  \label{eq:891}
  \nonumber j_{MAP} & = \arg \max_{j\in \mathcal H} Pr[\mathbf X(1,j)=1\big| \mathcal I_1, \mathcal I_2]\\
  \nonumber & \buildrel (a)\over = \arg \max_{j\in \mathcal H} Pr[\mathbf X(1,j)=1\big| \mathcal I_1]\\  
\nonumber & = \arg \max_{j\in \mathcal H} \sum_{\substack{\{i_1,...,i_{m-1}\}\subseteq \mathcal H\\j\not \in \{i_1,...,i_{m-1}\}}} Pr[\mathcal  G=\{j,i_1,...,i_{m-1}\}\big| \mathcal I_1]\\
  & \buildrel (b)\over = \arg \max_{j\in \mathcal H} \sum_{\substack{\{i_1,...,i_{m-1}\}\subseteq \mathcal H\\j\not \in \{i_1,...,i_{m-1}\}}} Pr[\mathcal I_1\big| \mathcal G=\{j,i_1,...,i_{m-1}\}],
\end{align}
where (a) follows since $\mathbf X(1,j)$ is independent of $\mathcal I_2$, and (b) is true due to the Bayes' rule, since all the $m$-tuples are equiprobable candidates for $\mathcal G$, because of the i.i.d.\ nature of the  columns of $\mathbf X$. We observe that if $j\in \mathcal G$, then $$\left|\mathbf y_{j}^{(1)}\right|_1\sim B(T_1, (1-p)(1-\epsilon)), \text{ and }\left|\mathbf y_{j}^{(2)}\right|_1\sim B(k-T_1,(1-p)(1-\epsilon)),$$ and if $j\in B$, then  
$$\left|\mathbf y_{j}^{(1)}\right|_1\sim B(T_1, p(1-\epsilon)),\text{ and } \left|\mathbf y_{j}^{(2)}\right|_1\sim B(k-T_1, p(1-\epsilon)).$$
It is also true that $\left\{ \left\{\left|y_j^{(1)}\right|_1\right\}_{j\in \mathcal G}, \left\{\left|y_j^{(2)}\right|_1\right\}_{j\in \mathcal G}, \left\{\left|y_j^{(1)}\right|_1\right\}_{j\in \mathcal B}, \left\{\left|y_j^{(2)}\right|_1\right\}_{j\in \mathcal B}\right\}$ are all independent of each other (conditioned on $\mathbf X(1,\mathcal H)$). Thus we have
\begin{align}
  \label{eq:892}
\nonumber   & Pr[\mathcal I_1\big| \mathcal G=\{j,i_1,...,i_{m-1}\}]\\
\nonumber \buildrel (a)\over= & Pr\left[\left\{(|\mathbf y_j^{(1)}|_1, |\mathbf y_j^{(2)}|_1)=(s_j^{(1)}, s_j^{(2)}): j\in \mathcal H\right\}\big|\mathcal G=\{j,i_1,...,i_{m-1}\}\right]\\
 \nonumber  = &\prod_{t\in \{j,i_1,...,i_{m-1}\}} {T_1 \choose s_t^{(1)}}{k-T_1 \choose s_t^{(2)}}\big((1-p)(1-\epsilon)\big)^{s_t^{(1)}+s_t^{(2)}} \big(1-(1-p)(1-\epsilon)\big)^{k-(s_t^{(1)}+s_t^{(2)})}\\
   &\hspace{1in}\cdot\prod_{\substack{l\not \in \{j,i_1,...,i_{m-1}\}\\ l \in \mathcal H}} {T_1 \choose s_l^{(1)}}{k-T_1 \choose s_l^{(2)}}\big(p(1-\epsilon)\big)^{s_l^{(1)}+s_l^{(2)}} \big(1-p(1-\epsilon)\big)^{k-(s_l^{(1)}+s_l^{(2)})},
\end{align}
where (a) follows due to the definition of $\mathcal I_1$.
Thus, for $j,j'\in \mathcal H$ such that $j, j'\not \in \{i_1,...,i_{m-1}\}$, we have
\begin{align}
  \label{eq:893}
  \nonumber & \frac{ Pr[\mathcal I_1\big| \mathcal G=\{j,i_1,...,i_{m-1}\}]}{ Pr[\mathcal I_1\big| \mathcal G=\{j',i_1,...,i_{m-1}\}]}\\
  \nonumber= &\frac{\big((1-p)(1-\epsilon)\big)^{s_j^{(1)}+s_j^{(2)}} \big(1-(1-p)(1-\epsilon)\big)^{-(s_j^{(1)}+s_j^{(2)})} \big(p(1-\epsilon)\big)^{s_{j'}^{(1)}+s_{j'}^{(2)}} \big(1-p(1-\epsilon)\big)^{-(s_{j'}^{(1)}+s_{j'}^{(2)})}}{\big((1-p)(1-\epsilon)\big)^{s_{j'}^{(1)}+s_{j'}^{(2)}} \big(1-(1-p)(1-\epsilon)\big)^{-(s_{j'}^{(1)}+s_{j'}^{(2)})} \big(p(1-\epsilon)\big)^{s_{j}^{(1)}+s_{j}^{(2)}} \big(1-p(1-\epsilon)\big)^{-(s_{j}^{(1)}+s_{j}^{(2)})}}\\
  = &\left(\frac{(1-p)}{p}\frac{(1-p(1-\epsilon))}{(1-(1-p)(1-\epsilon)}\right)^{s_j^{(1)}+s_j^{(2)}-(s_{j'}^{(1)}+s_{j'}^{(2)})}.
\end{align}
Since $p<1/2$, we now see that if $s_j^{(1)}+s_j^{(2)} \ge s_{j'}^{(1)}+s_{j'}^{(2)}$, then from \eqref{eq:893}
\begin{align}
\label{eq:894}Pr[\mathcal I_1\big| G=\{j,i_1,...,i_{m-1}\}] \ge Pr[\mathcal I_1\big| G=\{j',i_1,...,i_{m-1}\}].
\end{align}
From the above calculations, we also see that for $j,j'\in \mathcal H$ such that  $\{j,i_1,...,i_{m-1}\}=\{j',i_1',....i_{m-1}'\}$, we have 
\begin{align}
  \label{eq:895}
  \frac{ Pr[\mathcal I_1\big|\mathcal  G=\{j,i_1,...,i_{m-1}\}]}{ Pr[\mathcal I_1\big|\mathcal G=\{j',i_1',...,i_{m-1}'\}]} = 1.
\end{align}
Thus in \eqref{eq:891}, each term in the summation is maximized for the column $j$ with maximal $s_j^{(1)}+s_j^{(2)}$. Thus we have from \eqref{eq:891},
\begin{align}
  \label{eq:896}
  j_{MAP} = \arg \max _{j:\mathbf Y_k(1,j)=*} s_j^{(1)} +s_j^{(2)},
\end{align}
and hence is the same as choosing the column $j$ of $\mathbf Y_k$ with most number of 1's. Thus the probability of error for this MAP decoder is same as the error probability of the $\mathtt{PAF}$ algorithm for $T=k$. To be precise, we have now shown that
\begin{align}
  \label{eq:897}
  P_e^{MAP}= P_e\left[\mathtt{PAF}(k) \big| E_{1,n}^c\right].
\end{align}
Instead of using the MAP decoder, if we use the  decoder that chooses the column that maximizes $s_j^{(1)}$, then it's error probability is same as that of choosing the column $j$ of  $\mathbf Y_{T_1}$ with most number of 1's. To be precise, suppose we use the following sub-optimal decoder that chooses
$$j_{sub-optimal}^{(1)} :=\arg \max _{j:\mathbf Y_k(1,j)=*} s_j^{(1)}.$$
Then it's error probability is
\begin{align}
\label{eq:900}P_e^{sub-optimal, (1)}=P_e\left[\mathtt{PAF}(T_1)\big| E_{1,n}^c\right].
\end{align}
Similarly a different sub-optimal decoder that chooses
\begin{align}
\label{eq:901}j_{sub-optimal}^{(2)} :=\arg \max _{j:\mathbf Y_k(1,j)=*} s_j^{(1)}+s_j^{(2)}+s_j^{(b)},
\end{align}
has error probability 
$$P_e^{sub-optimal, (2)}=P_e\left[\mathtt{PAF}(T_2)\big|E_{1,n}^c\right].$$
Since MAP is a minimum error probability \cite[p. 8]{Poor1} decoder, and since $T_1 < k <T_2$,  (\ref{eq:897}), (\ref{eq:900}) and (\ref{eq:901}) together now imply that if $T\neq k$, then 
\begin{align}
  \label{eq:902}
  P_e\left[\mathtt{PAF}(T)\big|E_{1,n}^c\right] \ge P_e\left[\mathtt{PAF}(k)\big|E_{1,n}^c\right]. 
\end{align}
By observing that $P[E_{1,n}]=o(1)$ due to \eqref{eq:1}, we now obtain
\begin{align}
\label{eq:903} P_e\left[\mathtt{PAF}(T)\right] \ge P_e\left[\mathtt{PAF}(k)\right]+o(1). 
\end{align}
Using $\lim\inf_{n\rightarrow \infty}$ to the left hand side, and $\lim\sup_{n\rightarrow \infty}$ to the right hand side of \eqref{eq:903} implies the lemma.

\subsection{Proof of Lemma \ref{lemma:04}}
\label{proof:lemma:04}
To prove this lemma we show that $\forall i=2,3,...,n$, $s_{1i}$ is dominated by a binomial random variable  $\bar s_{1i}\sim B(n, (1-\epsilon)^2)$. The lemma will follow by upper bounding the upper tail of $\bar s_{1j}$. 

We first define another quantity to measure the overlap between two rows.
\begin{equation}
\label{eq:def:02}\bar s_{ij}:= \sum_{k=1}^n \mathbf 1\{y_{ik}\neq *\}\cdot \mathbf 1\{y_{jk}\neq *\}.
\end{equation}
From \ref{eq:def:01},  we see that $\bar s_{1j}\ge s_{1j}$ for all $j$.
We first lower bound the upper tail of $\bar s_{1j}$.
  We see that $\forall j=2,3,...,n$ we have $\bar s_{1j}\sim B(n, (1-\epsilon)^2)$. Hence 
  \begin{align*}
    Pr[\bar s_{1j} \ge t] &= \sum_{s=t}^n Pr[\bar s_{1j}=s]\\
    & = \sum_{s=t}^n {n \choose s} (1-\epsilon)^{2s}(1-(1-\epsilon)^2)^{n-s}\\
    & \buildrel (a)\over \le \sum_{s=t}^\infty (c^2n^{-2\beta})^s\\
    & \quad= \frac{c^{2t}n^{-2\beta t}}{1-c^2n^{-2\beta}}\\
    &\quad \buildrel (b) \over \le 2 c^{2t}n^{-2\beta t}, \text{ for large $n$},
 \end{align*}
 where (a) follows since ${n\choose s}\le n^s, 1-\epsilon=\frac{c}{n^\alpha}$, and $1-(1-\epsilon)^2 \le 1$, and (b) is true because $1-cn^{-2\beta t} \rightarrow 1$ with $n$.
Thus the probability that the overlap is more than $t$ for some row $=2,3,...,n$ is (by union bound)
$$Pr[\exists j, \bar s_{1j} > t]\le 2 (n -1)c^{2t}n^{-2\beta t} \rightarrow 0,$$ if $2\beta t >1$, i.e. if $t > \frac{1}{2\beta}$. Defining $t_{max}:=\lfloor \frac{1}{2\beta}\rfloor$ proves that w.h.p. for all $j=2,3,...,n$, $\bar s_{1j}\le t_{max}$. Since $\bar s_{1j}\ge s_{1j}$, we now have that w.h.p. for all $j=2,3,...,n$, $S_{1j}\le t_{max}$.

\subsection{Proof of lemma \ref{lemma:const_ratio}}
\label{proof:lemma:const_ratio}
To prove this lemma, we see that for $k > c_4n^{m(2\alpha -1)}\log r$, $N_j(m)$ are mixtures of Binomials with ``high'' mean, which lead to ``strong'' concentration around the mean. This implies that $\{N_j(m)\}_{j=1}^r$ are within constant factors of each other.  But, when $k \le c_4n^{m(2\alpha -1)}\log r$, we do not have ``strong'' concentration in general due to low mean, but we can suitably upper bound $N_1(m)$ and find a lower bound for the other $N_j(m)$'s. Finally when $k$ becomes much smaller ($=o(n^{m(2\alpha -1)})$), $N_1(m)=0$ w.h.p. Below, we see this in detail.

\begin{enumerate}
\item 
First we study the order of $N_{good}(m)$.
Let $L$ be the number of unerased entries of  row 1. We see that $L\sim B(n, 1-\epsilon)$, hence $\mathbb E[L]=cn^{1-\alpha}$ and w.h.p. $L = \Theta\left(n^{1-\alpha}\right)$, due to the Chernoff bound. Conditioned on the erasure sequence of row 1,  we see that $\forall j\in A_1$, $s_{1j}\sim B(L,1-\epsilon)$.  Let $\forall j\in A_1$,
\begin{align}
\label{eq:110}p_l(m):=Pr[s_{1j}=m \big| L=l]={l \choose m} (1-\epsilon)^m \epsilon^{l-m}.
\end{align}
Conditioned on $L=l$, every row $j\in A_1$ contributes to $N_{good}(m)$ independently with probability $p_l(m)$, implying $N_{good}(m)\sim B(k,p_l(m))$.  
  Now for $l=\Theta\left(n^{1-\alpha}\right)$,
\begin{align}
\nonumber \mathbb E[N_{good}(m)\big| L=l]&=kp_l(m)\\
\nonumber & = k {l\choose m} (1-\epsilon)^{m}\epsilon^{l-m}\\
\nonumber & \buildrel (a)\over = \Theta\left(\frac{k l^m} {n^{\alpha m}}\right)\\
\label{eq:ch09} & =\Theta\left(\frac{k}{n^{m(2\alpha-1)}}\right)= \Theta\left( \frac{k}{n^{2\beta m}}\right),
\end{align}
where (a) follows since  for a constant $m$ we have ${l\choose m} \sim l^m$, $1-\epsilon=c/n^\alpha$, and $\epsilon^{l-m} \rightarrow 1$ for $l=\Theta\left(n^{1-\alpha}\right)$. 
Since $k > c_4 n^{2\beta m} \log r$, conditioned on $L=l=\Theta(n^{1-\alpha})$, applying the  Chernoff bound \cite[Theorem 1.1]{Dubhashi1} on $N_{good}(m)$  we see that for any $\delta \in (0,1)$, w.h.p. 
\begin{align}
\label{eq:ch10} \mathbb E[N_{good}(m)|L=l](1-\delta)\le N_{good}(m) \le \mathbb E[N_{good}(m)|L=l](1+\delta),
\end{align}
We have already seen that w.h.p. $L = \Theta\left(n^{1-\alpha}\right)$. Thus \eqref{eq:ch09} and \eqref{eq:ch10} now imply that  w.h.p. $N_{good}(m)=\Theta\left(\frac{k}{n^{2\beta m}}\right)$.

Now we obtain a similar order bound for each of $N_i(m)$. Recall the definition of $C_j$ that it denotes the number of common column clusters of $\mathbf X$ between row cluster 1 and $j$. Then we see that  for $j=2,...,r$, $C_j \sim B(r,1/2)$, and thus  the Chernoff bound \cite[Theorem 1.1]{Dubhashi1} implies that $\forall j=2,...,r$, w.h.p.  $C_j = \Theta(r)$ as long as $r$ increases to infinity with $n$. 
For a row cluster $A_j$, 
let $Q_j$ be the number of  unerased entries  of row 1, restricted to these $C_j$ common column clusters. Conditioned on the value of $C_j=c_j$, we see that $Q_j\sim B(c_j k, 1-\epsilon)$, implying $\mathbb E[Q_j|C_j=c_j]=c_jk(1-\epsilon) $. Since $n=kr$, we have for  $c_j=\Theta(r)$, 
$$\mathbb E[Q_j|C_j=c_j]=\Theta(n)(1-\epsilon)=\Theta(n^{1-\alpha}),$$
 hence using the Chernoff bound we see that for $\delta\in (0,1)$, conditioned on $C_j=c_j$,  w.h.p.
\begin{align}
  \label{eq:ch12}
  \mathbb E[Q_j|C_j=c_j](1-\delta) \le Q_j \le \mathbb E[Q_j|C_j=c_j](1+\delta).
\end{align}
Since w.h.p. $C_j=\Theta(r)$,  \eqref{eq:ch12} now implies that w.h.p. $Q_j =\Theta\left(n^{1-\alpha}\right)$ .

Let $\hat s_i$ denote the number of commonly sampled entries of row 1 and row $i\in A_j$ within these $C_j$ common column clusters. Then conditioned on $Q_j=q$, we see that $\hat s_i\sim B(q, 1-\epsilon)$.  Thus 
$$Pr[\hat s_i = m \big| Q_j=q]= {q\choose m} (1-\epsilon)^m \epsilon^{q-m}=p_q(m),$$
where $p_q(m)$ is as defined in \eqref{eq:110}. We see that conditioned on $Q_j=q$, each row $i\in A_j$ overlaps with row 1 at $m$ entries independently with probability $p_q(m)$, i.e.,  $N_j(m)\sim B(k, p_q(m))$ and thus, for $q=\Theta\left(n^{1-\alpha}\right)$, we have
\begin{align}
\nonumber  \mathbb E[N_j(m)\big|Q_j=q] & = kp_q(m)\\
\nonumber   &  = k {q\choose m} (1-\epsilon)^m \epsilon^{q-m}\\
\nonumber& \buildrel (b) \over = \Theta\left(\frac{k q^m}{n^{m\alpha}}\right)\\
\label{eq:new} & = \Theta\left(\frac{k}{n^{m(2\alpha-1)}}\right)  = \Theta\left(\frac{k}{n^{2\beta m}}\right),
\end{align}
where (b) follows since for a constant $m$ we have ${q\choose m} = \Theta(q^m)$, $1-\epsilon=c/n^\alpha$, and $\epsilon^{q-m} \rightarrow 1$ for $q=\Theta\left(n^{1-\alpha}\right)$. 
Since $k > c_4 n^{2\beta m} \log r$ for a large enough constant $c_4$, conditioned on $Q_j=q_j=\Theta(n^{1-\alpha})$, the Chernoff bound applied to $N_j(m)$  along with an union bound gives that w.h.p. $\forall j=2,3,...,r$ we have
\begin{align}
\label{eq:ch11}  \mathbb E[N_j(m)|Q_j=q_j](1-\delta)\le N_j(m) \le \mathbb E[N_j(m)|Q_j=q_j](1+\delta).
\end{align} 
As we have already seen that $Q_j=\Theta\left(n^{1-\alpha}\right)$ , \eqref{eq:new} and  \eqref{eq:ch11} now imply that w.h.p. 
$$\{N_j(m)\}_{j=2}^r =\Theta\left(\frac{k}{n^{2\beta m}}\right).$$ This along with the previous observation that w.h.p.  $N_{good}(m)=\Theta\left(\frac{k}{n^{2\beta m}}\right)$, proves the first part of the lemma.

\item 
Before we start proving the second part of the lemma, we need a bound on the upper binomial tail with small mean.
\begin{lemma}[{Tail of a binomial\cite[p. 23]{Dubhashi1}}]
\label{lemma:low_mean}
  Suppose $X\sim B(n,p)$ such that $\mathbb E[X]=np$. For $t > 2e\mathbb E[X]$, we have
\[Pr[X > t] \le 2^{-t}.\]
\end{lemma}

In the proof of the first part, we have seen that conditioned on  $L=l$, we have $N_{good}(m)\sim B(k,p_l(m))$, and for $l=\Theta\left(n^{1-\alpha}\right)$ we have 
$$\mathbb E[N_{good}(m)|L=l]=\Theta\left(\frac{k}{n^{2\beta m}}\right)=O(\log r),$$ where the last equality follows since $k < c_4n^{2\beta m}\log r$.  Thus, for a large enough constant $c'$ and for $t\ge c'\log r$, we have
 $$ Pr[N_{good}(m) > t] < 2^{-t},$$
which implies that w.h.p. $N_{good}(m)=O(\log r)$.  We have also seen in the proof of the first part, that for $j=2,3,...,r$, conditioned on $Q_j=q$, $N_j(m)\sim B(k,p_q(m))$ and for $q=\Theta(n^{1-\alpha})$ we have
\begin{align}
\nonumber E\left[N_j(m)|Q_j=q\right]=kp_q(m)= \Theta\left(\frac{k}{n^{2\beta m}}\right) =O(\log r),
\end{align}
where the last equality follows since $k < c_4n^{2\beta m}\log r$. Thus Lemma \ref{lemma:low_mean} together with an union bound implies that w.h.p.
$$\{N_j(m)\}_{j=2}^r =O(\log r).$$

Now we want to lower bound  $N_j(m)$ for $j >1$. Recall that for $j=2,3,...,r$, conditioned on $Q_j=q$, $N_j(m)\sim B(k, p_q(m))$ and for $q=\Theta(n^{1-\alpha})$ we have
 \begin{align}
\label{eq:330}E\left[N_j(m)|Q_j=q\right]=kp_q(m)= \Theta\left(\frac{k}{n^{2\beta m}}\right) > c_5', 
\end{align}
for a constant $c_5' > 0$, where the last inequality is true because $k > c_5 n^{2\beta m}$. Thus, for $q=\Theta(n^{1-\alpha})$  we have
\begin{align*}
  p_0^{(q)}&:=Pr[N_j(m)=0|Q_j=q]\\
  & \buildrel (a)\over=(1-p_q(m))^k\\
  & \le e^{-kp_q(m)}\\
  &\buildrel (b) \over < e^{-c_5'}\\
  & <1,
\end{align*}
where (a) is true since conditioned on $Q_j=q$, $N_j(m)\sim B(k,p_q(m))$, and (b) follows due to \eqref{eq:330}. 
Now we observe that conditioned on the values of $\{Q_2, Q_3, ...,Q_r\}$, $\{N_j(m)\}_{j=2}^r$ are independent random variables. Let $S$ denote the set of row clusters $j\in \{2,3, ..., r\}$ such that $N_j(m)\ge 1$.  Conditioned on the values $Q_j=q_j$ for $j=2,...,r$, we see that 
$$|S|=\sum_{j=2}^r \mathbf 1_j,$$ 
where $\{\mathbf 1_j\}_{j=2}^r$ are independent binary random variables with $Pr[\mathbf 1_j = 0]=p_0^{(q_j)}.$
 In the first part of the proof, we have seen that w.h.p. for all $j=2,...,r$, $Q_j=\Theta\left(n^{1-\alpha}\right)$.  This along with a Chernoff bound on $|S|$ implies that for any $\delta \in (0,1)$,  w.h.p. 
$$|S|> (r-1)(1-e^{-c_5'})(1-\delta).$$  In other words there exists a subset $S$ of $[r]\backslash  \{1\}$ such that $|S|=\Omega(r)$ and for $\forall j\in S, N_j(m)\ge 1$. 

\item 
We have already seen in the first part of the proof that w.h.p. $\mathbb E[N_{good}(m)]=\Theta\left(\frac{k}{n^{2\beta m}}\right)$.  Since $k=o(n^{2\beta m})$, we have $\mathbb E[N_{good}(m)]\rightarrow 0$. Thus 
$$Pr[N_{good}(m) >0] \rightarrow 0,$$ since for a positive integer valued random variable $X$, $$\mathbb E[X]=\sum_{i=0}^\infty Pr[X > 0].$$ 
\end{enumerate}

\subsection{Proof of Corollary \ref{cor:const_ratio}}
\label{proof:cor:const_ratio}
\begin{enumerate}
  \item The first part follows by observing that for $k > c_4n^{m(2\alpha -1)}\log r$, w.h.p.
    \begin{align*}
      N_j(m^+)=\sum_{t=m}^{t_{max}}N_j(t) & \buildrel (a)\over=\sum_{t=m}^{t_{max}} \Theta\left(\frac{k}{n^{t(2\alpha-1)}}\right)\\
      & = \Theta\left(\frac{k}{n^{m(2\alpha-1)}}\right),
    \end{align*}
    where (a) follows from the first part of  Lemma \ref{lemma:const_ratio}.

  \item For the second part, suppose $c_5n^{m(2\alpha -1)} \le k \le c_4n^{m(2\alpha -1)}\log r$.  Then for $t\ge m+1$ we have $k=o(n^{t(2\alpha -1)})$. Thus w.h.p.
    \begin{align*}
      N_j(m^+) &=N_{j}(m)+\sum_{t=m+1}^{t_{max}} N_{j}(t)\\
      & \buildrel (b) \over =O(\log r) +\sum_{t={m+1}}^{t_{max}}o(1) = O(\log r),
    \end{align*}
    where (b) is due to the second part of Lemma \ref{lemma:const_ratio}. The fact that there exists a subset $S$ of $[r]\backslash \{1\}$ with $|S|=\Omega(r)$ such that for $j\in S$ we have $N_j(m^+)\ge 1$, follows immediately from the second part of Lemma \ref{lemma:const_ratio}.

  \item For $k=o(n^{m(2\alpha -1)})$, we have  w.h.p. 
    \begin{align*}
      N_{good}(m^+) &=\sum_{t=m}^{t_{max}} N_{good}(t) = \buildrel (c) \over = 0,
    \end{align*}
    where (c) follows from the third part of Lemma \ref{lemma:const_ratio}.
  \end{enumerate}

\subsection{Proof of Lemma \ref{lemma:topT}}
\label{proof:lemma:topT}
We prove this lemma by using similar steps as used in proving Lemma \ref{lemma:const_ratio}, the main difference is that we need a tail bound for hyper-geometric random variables, unlike the Chernoff bound for i.i.d. random variables used in proving Lemma \ref{lemma:const_ratio}.

To begin with, we observe from \eqref{eq:325} and Lemma \ref{lemma:const_ratio} that 

\noindent
{\bf Obs.1)}  If $k>c_4n^{m_0(2\alpha -1)}\log r$, then we have w.h.p. $\{N_j(m_0)\}_{j=2}^r=\Theta(N_1(m_0))$, implying $\{\mathbb E[\xi_j]\}_{j=2}^r=\Theta(\mathbb E[\xi_1])$.

\noindent
{\bf Obs.2)} If there is a positive constant $c_5>0$, such that $c_5n^{m_0(2\alpha -1)}<k\le c_4n^{m_0(2\alpha -1)}\log r$, then w.h.p. $\mathbb E[\xi_1]=O(\log r)$.

\noindent
{\bf Obs.3)} If $k=o(n^{m_0(2\alpha -1)})$, then w.h.p. $\mathbb E[\xi_1]=0$.

We break down the proof into various cases for different values of $T$ and $k$. As in \eqref{eq:323}, suppose $m_0$ is a positive integer such that
$$N((m_0+1)^+) < T \le N(m_0^+),$$
and $c_4>0$ is a large positive constant (same as the constant $c_4$ defined in Lemma \ref{lemma:const_ratio}).

\noindent
 {\bf Case 1 ($k > c_4n^{(m_0+1)(2\alpha -1)}\log r$):}  Corollary \ref{cor:const_ratio}  implies that w.h.p. 
 \begin{align}
\label{eq:511}\{N_j((m_0+1)^+)\}_{j=1}^r=\Theta\left(\frac{k}{n^{(m_0+1)(2\alpha -1)}}\right) =\Omega(\log r),
\end{align}
and Theorem \ref{lemma:const_ratio} implies that w.h.p.
$$\{N_j(m_0)\}_{j=1}^r=\Theta\left(\frac{k}{n^{(m_0)(2\alpha -1)}}\right).$$
Also recall the definition of the hyper-geometric random variable $\xi_j$ from \eqref{eq:324}. Suppose $c_6$ is a large enough positive constant. We consider two possible cases.
\begin{enumerate}
\item Suppose $\min_j \mathbb E[\xi_j]> c_6\log r$. Since $\{\xi_j\}$ are hyper-geometric random variables, from the hyper-geometric tail bound (Corollary \ref{cor:tail}, Appendix \ref{app:hyper}) used together with an union bound, it follows that w.h.p. 
$$\{\xi_j\}_{j=1}^r=\Theta(\mathbb E[\xi_1])=\Omega(\log r),$$
and this together with \eqref{eq:324} and \eqref{eq:511} implies that  w.h.p. 
$\{T_j\}_{j=2}^r=\Omega(T_1).$ Thus there exists a positive integer $d$ such that w.h.p. for $j=2,3,...,r$, we have $T_j\ge dT_1$ for large enough $n$. This implies ($C_1$). 
\item Now suppose $\min_j \mathbb E[\xi_j] \le c_6 \log r$.  From Obs.1), we see that for different values of $j$, $\mathbb E[\xi_j]$ are within a constant factor of each other.  Thus we have $\mathbb E[\xi_1]=O(\log r)$. Then Corollary \ref{cor:low_mean_hyper} (see Appendix \ref{app:hyper}) implies that w.h.p. 
$$\xi_1=O(\log r).$$ 
This together with \eqref{eq:324} and \eqref{eq:511} implies that  w.h.p. $\{T_j\}_{j=2}^r=\Omega(T_1).$ This implies ($C_2$).
\end{enumerate}

\noindent
{\bf Case 2 ($\exists c_5 >0$, $c_5 n^{(m_0+1) (2\alpha -1)}  < k \le c_4n^{(m_0+1)(2\alpha -1)}\log r$):} Corollary \ref{cor:const_ratio} implies that w.h.p. 
$$N_1((m_0+1)^+) =O(\log r),$$
and there is a subset $S$ of $[r]\backslash \{1\}$ with $|S|=\Theta(r)$ such that for $j\in S$,  $$N_j((m_0+1)^+)\ge 1.$$ 
We now see from Obs.1) that $\mathbb E[\xi_1]=O(\log r)$, which together with Corollary\ref{cor:low_mean_hyper} in Appendix \ref{app:hyper} implies that w.h.p. 
$$\xi_1=O(\log r).$$
Thus  \eqref{eq:324}  now implies that  w.h.p. $T_1=O(\log r)$, and there is a subset $S$ of $[r]\backslash \{1\}$ with $|S|=\Theta(r)$ such that for $j\in S$,  $T_j \ge 1$. This implies ($C_2$).

\noindent
{\bf Case 3 ($c_4 n^{m_0(2\alpha -1)}\log r < k =o(n^{(m_0+1)(2\alpha -1)})$):} In this regime, Corollary \ref{cor:const_ratio} implies that w.h.p.
$$N_1((m_0+1)^+)=0.$$
Depending on the value of $\mathbb E[\xi_j]$, we now consider three possible cases.  Suppose $c_6$ is a large enough positive constant.
\begin{enumerate}
\item Suppose $\min_j \mathbb E[\xi_j]> c_6\log r$. Then from the hyper-geometric tail bound (Corollary \ref{cor:tail}, Appendix \ref{app:hyper}) used together with an union bound, it follows that w.h.p. 
$$\{\xi_j\}_{j=1}^r=\Theta(\mathbb E[\xi_1])=\Omega(\log r),$$
and this together with \eqref{eq:324} implies that  w.h.p. 
$\{T_j\}_{j=2}^r=\Omega(T_1).$ In other words, there exists a positive integer $d$ such that w.h.p. for $j=2,3,...,r$, we have $T_j\ge dT_1$ for large enough $n$, implying ($C_1$).

\item 
Now suppose $\exists c_5$, $c_5>0$, such that $c_5<\min_j \mathbb E[\xi_j] \le c_6 \log r$. Using Obs.1), this actually implies $\{\mathbb E[\xi_j]\}_{j=1}^r=O(\log r).$ Then the hyper-geometric tail bound (Corollary \ref{cor:low_mean_hyper} , Appendix \ref{app:hyper}), together with an union bound implies that w.h.p. 
\begin{align}
\label{eq:601}\{\xi_j\}_{j=1}^r=O(\log r).
\end{align}
Suppose $S':=\{i\in [r]: \xi_j \ge 1\}$. Since we have seen in \eqref{eq:325:1} that $\sum_{j=1}^r \xi_j =T-1-N((m_0+1)^+)$, \eqref{eq:601} implies that w.h.p.
\begin{align}
  \label{eq:602}
  |S'|=\Omega\left(\frac{T-1-N((m_0+1)^+)}{\log r}\right).
\end{align}
Since $k> c_4 n^{m_0(2\alpha -1)}\log r$,  using Lemma \ref{lemma:const_ratio} we see that $\{N_j(m_0)\}_{j=2}^r=\Theta(N_1(m_0))$,  implying
 $$\{N_j(m_0)\}_{j=1}^r=\Theta(N(m_0)/r).$$
 This observation together with \eqref{eq:325} implies that 
\begin{align}
  \label{eq:631}
\nonumber   \mathbb E[\xi_1] &=\frac{N_1(m_0)}{N(m_0)}(T-1-N((m_0+1)^+)\\
& = \Theta(1/r) (T-1-N((m_0+1)^+).
\end{align}
Since $\mathbb E[\xi_1]>c_5$, \eqref{eq:631}  now implies that
$$T-1-N((m_0+1)^+)=\Omega(r).$$
Thus from \eqref{eq:602} we see that w.h.p.
$$|S'|=\Omega(r/\log r).$$ 
Now using \eqref{eq:324} we see that w.h.p. $T_1=O(\log r)$, and for $j\in S'$, $T_j\ge 1$, implying ($C_2$).
\item If $\min_j \mathbb E[\xi_j]\rightarrow 0$, then using Obs.1), we see that $\mathbb E[\xi_1] \rightarrow 0$. This implies that w.h.p. $\xi_1=0$, since for a non-negative integer valued random variable $X$, $\mathbb E[X]=\sum_{i=0}^\infty Pr[X>i]$. As we have already observed that w.h.p. $N_1((m_0+1)^+)=0$, \eqref{eq:324} now implies that w.h.p. $T_1=0$. This implies ($C_3$).
\end{enumerate}

\noindent
{\bf Case 4 ($k\le c_4 n^{m_0(2\alpha -1)}\log r$):} In this regime, Corollary \ref{cor:const_ratio} implies that w.h.p.
$$N_1((m_0+1)^+)=0,$$
since $k=o\left(n^{(m_0+1)(2\alpha -1)}\right)$. If $k=o(n^{m_0(2\alpha -1)})$, then Lemma \ref{lemma:const_ratio} implies that w.h.p.
$$N_1(m_0)=0,$$
implying w.h.p. $\xi_1=0$, which together with \eqref{eq:324} implies that w.h.p. $T_1=0$, and hence ($C_3$). Thus we now assume that there is a constant $c_5>0$, such that $k > c_5n^{m_0(2\alpha-1)}$.
Using Lemma \ref{lemma:const_ratio} we see that 
$$\{N_j(m_0)\}_{j=1}^r =O(\log r),$$
implying 
\begin{align}
\label{eq:610}\{\mathbb E[\xi_j]\}_{j=1}^r=O(\log r),
\end{align}
due to (\ref{eq:325}). Depending on the value of $\mathbb E[\xi_1]$, we now consider two possible cases.  Suppose $c_6$ is a large enough positive constant.
\begin{enumerate}
\item Suppose $\exists c_6$, $c_6>0$ such that $c_6<\mathbb E[\xi_1] \le c_6 \log r$.  Using \eqref{eq:610} and the hyper-geometric tail bound (Corollary \ref{cor:low_mean_hyper}, Appendix \ref{app:hyper}), together with an union bound implies that w.h.p. 
\begin{align}
\label{eq:604}\xi_1=O(\log r).
\end{align}
As in the second part of Case 3, suppose $S':=\{i\in [r]: \xi_j \ge 1\}$. Since we have seen in \eqref{eq:325:1} that $\sum_{j=1}^r \xi_j =T-1-N((m_0+1)^+)$, \eqref{eq:604} implies that w.h.p.
\begin{align}
  \label{eq:605}
  |S'|=\Omega\left(\frac{T-1-N((m_0+1)^+)}{\log r}\right).
\end{align}
Since $$c_5 n^{m_0(2\alpha-1)} <k\le c_4 n^{m_0(2\alpha -1)}\log r,$$
  using Lemma \ref{lemma:const_ratio} we see that $N_1(m_0)=O(\log r)$ and there is a subset $S$ of $[r]\backslash \{1\}$ with $|S|=\Theta(r)$, such that for $j\in S$, $N_j(m_0)\ge 1$. Thus we have  $N(m_0)=\Omega(r)$.
Using this observation together with \eqref{eq:325} we see that 
\begin{align}
  \nonumber \mathbb E[\xi_1] & = \frac{N_1(m_0)}{N(m_0)}(T -1-  N((m_0+1)^+))\\
  \label{eq:645}& =O(\log r/r) (T -1- N((m_0+1)^+)).
\end{align}
Since $\mathbb E[\xi_1]>c_5,$ \eqref{eq:645} now implies that
$$T-1-N((m_0+1)^+)=\Omega(r/\log r).$$
Thus from \eqref{eq:605} we see that w.h.p.
$$|S'|=\Omega(r/\log^2 r).$$ 
Now using \eqref{eq:324} we see that w.h.p. $T_1=O(\log r)$, and for $j\in S'$, $T_j\ge 1$.  This implies ($C_2$).

\item Now suppose $\mathbb E[\xi_1]\rightarrow 0$.  This implies that w.h.p. $\xi_1=0$, since for a non-negative integer valued random variable $X$, $\mathbb E[X]=\sum_{i=0}^\infty Pr[X>i]$. As we have already observed that w.h.p. $N_1((m_0+1)^+)=0$, \eqref{eq:324} now implies that w.h.p. $T_1=0$. This implies ($C_3$).
\end{enumerate}

\subsection{Proof of Lemma \ref{lemma:maj}}
\label{proof:lemma:maj}
To prove this lemma, 
we consider a ``new'' estimation problem and consider two different estimators, the first of which is a maximum aposterior probability (MAP) estimator having probability of error equal to the right hand side (RHS) of Lemma \ref{lemma:maj}; whereas the second estimator is a sub-optimal one and has probability of error equal to the left hand side (LHS) of Lemma \ref{lemma:maj}. Since MAP estimator minimizes probability of error over all estimators \cite{Poor1}, this would prove the lemma.

By increasing the number of good neighbors from $T_1$ to $u_n+1$ (recall that  $u_n$ is the smallest multiple of $d$ not less than $T_1$), 
we  increase the number of 1's in the columns $j$ with $\mathbf X(1,j)=1$, and do not change the number of 1's in the columns $j$ with $\mathbf X(1,j)=0$. Thus this reduces  the probability of error for  majority decoding  on $\mathbf Y_T$. Thus to prove the lower bound on $P_e^{maj}[\mathbf Y_T]$, we  assume  without loss of generality that $T_1=u_n$.

For every row cluster $A_i, i\ge 2$ of $\mathbf Y_T$, suppose $A_i^{(1)}$ represents the first $l_n$ rows in that cluster, and $A_i^{(2)}$ represents the rest of the $T_i - l_n$ rows.  Consider the following estimation problem, where we do not get to observe $\mathbf Y_T$. Instead we observe the following two random variables.
\begin{itemize}
\item For all columns $j$ such that $\mathbf Y^{(e)}(1,j)=*$, we observe the corresponding column sums, i.e., we observe $|\mathbf Y^{(e)}(:,j)|_1=t_j$. Let $\mathcal I_1$ denote the collection of these observed random variables.
\item We also observe the column sums of $\mathbf Y_T$ restricted to the second part of the row clusters. To make this precise, let $\mathbf y_j$ denote $j$th column of $\mathbf Y_T$, restricted to $\cup_{i=2}^r A_i^{(2)}$. Then we observe $s_j:=|\mathbf y_j|_1$. Let $\mathcal I_2$ denote the collection of these observed random variables.
\end{itemize}
Upon observing $\mathcal I_1$ and $\mathcal I_2$, we want to find a column $j$ such that $\mathbf X^{(e)}(1,j)=1$. First we consider the MAP estimator for this problem, which selects a column $j_{MAP}$ satisfying
\begin{align}
\nonumber  j_{MAP}& :=\arg\max_{j: \mathbf Y^{(e)}(1,j)=*} Pr[\mathbf X^{(e)}(1,j)=1|\mathcal I_1, \mathcal I_2]\\
\nonumber  & \buildrel (a)\over = \arg\max_{j: \mathbf Y^{(e)}(1,j)=*} Pr[\mathbf X^{(e)}(1,j)=1|\mathcal I_1]\\
\label{eq:613}  & \buildrel (b) \over = \arg\max_{j: \mathbf Y^{(e)}(1,j)=*} Pr[\mathbf X^{(e)}(1,j)=1\big | |\mathbf Y^{(e)}(:,j)|_1=t_j],
\end{align}
where (a) follows since $\mathbf X^{(e)}(1,j)$ is independent of $\mathcal I_2$, as $\mathcal I_2$ contains information only about the bad row clusters, and (b) is true because 
$$\mathbf X^{(e)}(1,j)\longrightarrow |\mathbf Y^{(e)}(:,j)|_1 \longrightarrow \mathcal I_1.$$
We now state a lemma that will help in simplifying the above expression for $j_{MAP}$.
\begin{lemma}
  \label{lemma:increasing}
 $ Pr[\mathbf X^{(e)}(1,j)=1\big | |\mathbf Y^{(e)}(:,j)|_1=t_j]$ is an increasing function of $t_j$.
\end{lemma}
\begin{proof}
First we observe that $|\mathbf Y^{(e)}(:,j)|_1$ is a multiple of $l_n$, where $l_n$ is the size of bad row clusters of $\mathbf Y^{(e)}$. Thus $t_j=m_jl_n$ for some positive integer $m_j$.  We also see that  $u_n/l_n=d$ (where $d$ is as in Lemma \ref{lemma:topT}). 
  Let $$p_1:=Pr[\mathbf X^{(e)}(1,j)=1\big | |\mathbf Y^{(e)}(:,j)|_1=t_j],$$ and $p_0:=1-p_1$. Then
  \begin{align*}
    \frac{p_0}{p_1} &= \frac{Pr[\mathbf X^{(e)}(1,j)=0\big | |\mathbf Y^{(e)}(:,j)|_1=t_j]}{Pr[\mathbf X^{(e)}(1,j)=1\big | |\mathbf Y^{(e)}(:,j)|_1=t_j]}\\
    & \buildrel (a)\over = \frac{Pr[|\mathbf Y^{(e)}(:,j)|_1=t_j\big |\mathbf X^{(e)}(1,j)=0 ]}{Pr[|\mathbf Y^{(e)}(:,j)|_1=t_j\big |\mathbf X^{(e)}(1,j)=1 ]}\\
    & \buildrel (b) \over= \frac{{r-1 \choose m_j}}{{r-1 \choose m_j-d}}\\
    & = \frac {(m_j-d)! (r-1-m_j+d)!}{m_j! (r-1-m_j)!}\\
    & = \frac{(m_j-d+1)(m_j-d+2)\cdots m_j}{(r-m_j)(r-m_j+1)\cdots (r-1-m_j+d)}\\
    & = \frac{1}{\left(\frac{r}{m_j}-1\right)\left( \frac{r}{m_j-1} -1\right)\cdots \left( \frac{r}{m_j-d+1}-1\right)}.
  \end{align*}
where (a) is due to the Bayes' expansion and the observation that $Pr[\mathbf X^{(e)}(1,j)=1]=1/2$, (b) is true since for $j$ such that $\mathbf Y^{(e)}(1,j)=*$,  $|\mathbf Y^{(e)}(:,j)|_1\sim l_n\times B(r, 1/2).$ Thus $p_0/p_1$ is clearly a decreasing function of $m_j$.  But
$$\frac{p_0}{p_1}=\frac{1}{p_1}-1,$$
implying that $p_1$ is an increasing function of $m_j$. Since $t_j=m_jl_n$, we now see that $p_1$ is an increasing function of $t_j$.
\end{proof}
Using this lemma, \eqref{eq:613} now becomes
\begin{align*}
  j_{MAP}&= \arg\max_{j: \mathbf Y^{(e)}(1,j)=*}  |\mathbf Y^{(e)}(:,j)|_1\\
  &= j_{maj}(\mathbf Y^{(e)}).
\end{align*}
In other words, majority decoding is same as the MAP estimator for the above estimation problem. Now we consider a different (sub-optimal) estimator for the same problem.
Suppose 
\begin{align}
  \label{eq:614}
  \tilde j_{maj}:=\arg\max_{j:\mathbf Y^{(e)}(1,j)=*} B(|\mathbf Y^{(e)}(:,j)|_1, 1-\epsilon) + s_j,
\end{align}
where we recall that $s_j$ denotes the number of 1's in the $j$th column of $\mathbf Y_T$, restricted to $\cup_{i=2}^r A_i^{(2)}$. Let the corresponding probability of error be $\tilde P_e[\mathcal I]$.  We observe that the probability law of $B(|\mathbf Y^{(e)}(:,j)|_1+s_j$ is same as that of$|\mathbf Y_T(:,j)|$. Thus we have 
$$\tilde P_e[\mathcal I] = P_e^{maj}[\mathbf Y_T],$$
and this together with the fact that MAP estimator minimizes the probability of error \cite[p. 8]{Poor1},  implies that 
$P_e^{maj}[\mathbf Y_T] \ge P_e^{maj}[\mathbf Y^{(e)}].$

\subsection{Proof of Lemma \ref{lemma:maj_unif}}
\label{proof:lemma:maj_unif}

\noindent
{\bf Conditioned on $(C_1)$:}
We first condition on the event $(C_1)$. In this case, $\mathbf a_e$ is an $r$-length vectors with $\mathbf a_e(1)=u_n+1=d\left\lceil \frac{T_1}{d}\right\rceil+1$, and for $j=2,3,...,r$, $\mathbf a_e(j)=\left\lceil \frac{T_1}{d}\right\rceil=:l_1$. 
Let 
$$p_1:=Pr[\mathbf X^{(e)}(1,j)=1\big|j_{maj}(\mathbf Y^{(e)})=j],$$
and $$p_0:=1-p_1=Pr[\mathbf X^{(e)}(1,j)=0\big|j_{maj}(\mathbf Y^{(e)})=j].$$
By computing the ration $p_1/p_0$ using the Bayes' rule, it can be shown that the majority estimator is not worse than a random estimator, i.e., $p_1\ge 1/2$ and $p_0\le 1/2$.
For a column $j$ such that $\mathbf Y^{(e)}(1,j)=*$, conditioned on $\mathbf X^{(e)}(1,j)=1$, we have $|\mathbf Y^{(e)}(:,j)|_1 =l_1(d +\psi_j)$, where $\psi_j \sim B(r-1, 1/2)$. Thus  the Chernoff bound implies that for some $\delta_r> 0 $ with $\delta_r
 =o(r)$, w.h.p.
\begin{align}
\label{eq:conc}
|\mathbf Y^{(e)}(:,j)|_1\in \left[l_1\left(\frac{r}{2}-\delta_r\right), l_1\left(\frac{r}{2}+\delta_r\right)\right].
\end{align}
Let $A$ denote the interval $\left[\frac{r}{2}-\delta_r, \frac{r}{2}+\delta_r\right]$. Then by observing that $|\mathbf Y^{(e)}(:,j)|_1$ is a multiple of $l_1$,we see that
\begin{align}
  \nonumber p_1 & = \sum_{m=0}^r Pr[\mathbf X^{(e)}(1,j)=1, |\mathbf Y^{(e)}(:,j)|_1=l_1m\big| j_{maj}(\mathbf Y^{(e)})=j]\\
\nonumber
&\buildrel (a)\over = \sum_{m=0}^r Pr[\mathbf Y^{(e)}(:,j)|_1=l_1m \big | j_{maj}(\mathbf Y^{(e)})=j]  \cdot Pr[\mathbf X^{(e)}(1,j)=1\big | |\mathbf Y^{(e)}(:,j)|_1=l_1m] \\
\nonumber &\buildrel (b) \over = \sum_{m\in A} Pr[\mathbf Y^{(e)}(:,j)|_1=l_1m \big | j_{maj}(\mathbf Y^{(e)})=j]  \cdot Pr[\mathbf X^{(e)}(1,j)=1\big | |\mathbf Y^{(e)}(:,j)|_1=l_1m] +o(1)\\
\label{eq:p1}&\buildrel (c) \over \doteq \sum_{m\in A} Pr[\mathbf Y^{(e)}(:,j)|_1=l_1m \big | j_{maj}(\mathbf Y^{(e)})=j]  \cdot Pr[\mathbf X^{(e)}(1,j)=1\big | |\mathbf Y^{(e)}(:,j)|_1=l_1m] .
\end{align}
where (a) is true because of the Markov relation
$$\mathbf X^{(e)}(1,j)\longrightarrow |\mathbf Y^{(e)}(:,j)|_1 \longrightarrow \{j_{maj}=j\},$$
(b) follows due to \eqref{eq:conc}, and (c) is true since $p_1 >1/2$. Similarly we obtain
\begin{align}
  \label{eq:p0}
  p_0= \sum_{m\in A} Pr[\mathbf Y^{(e)}(:,j)|_1=l_1m \big | j_{maj}(\mathbf Y^{(e)})=j]  \cdot Pr[\mathbf X^{(e)}(1,j)=0\big | |\mathbf Y^{(e)}(:,j)|_1=l_1m] +o(1),
\end{align}
implying 
\begin{align}
  \label{eq:633}
 &\mathtt{ratio\_mean} :=\frac{P_e^{maj}(\mathbf Y^{(e)})}{1-P_e^{maj}(\mathbf Y^{(e)})}=\frac{p_0}{p_1}\\
\nonumber   \doteq  &\frac{\sum_{m\in A} Pr[\mathbf Y^{(e)}(:,j)|_1=l_1m \big | j_{maj}(\mathbf Y^{(e)})=j]  \cdot Pr[\mathbf X^{(e)}(1,j)=0\big | |\mathbf Y^{(e)}(:,j)|_1=l_1m] +o(1) }{\sum_{m\in A} Pr[\mathbf Y^{(e)}(:,j)|_1=l_1m \big | j_{maj}(\mathbf Y^{(e)})=j]  \cdot Pr[\mathbf X^{(e)}(1,j)=1\big | |\mathbf Y^{(e)}(:,j)|_1=l_1m]}\\
\nonumber   =  &\frac{\sum_{m\in A} Pr[\mathbf Y^{(e)}(:,j)|_1=l_1m \big | j_{maj}(\mathbf Y^{(e)})=j]  \cdot Pr[\mathbf X^{(e)}(1,j)=0\big | |\mathbf Y^{(e)}(:,j)|_1=l_1m] }{\sum_{m\in A} Pr[\mathbf Y^{(e)}(:,j)|_1=l_1m \big | j_{maj}(\mathbf Y^{(e)})=j]  \cdot Pr[\mathbf X^{(e)}(1,j)=1\big | |\mathbf Y^{(e)}(:,j)|_1=l_1m]}+o(1).
\end{align}
We now have
\begin{align}
  \label{eq:634}
  \texttt{ratio\_mean}\ge \texttt{ratio\_min}:= \min_{m\in A} \texttt{ratio}(m)+o(1),
\end{align}
where
\begin{align}
  \label{eq:635}
\nonumber \texttt{ratio}(m) &:=  \frac{Pr[\mathbf X^{(e)}(1,j)=0\big | |\mathbf Y^{(e)}(:,j)|_1=l_1m] }{Pr[\mathbf X^{(e)}(1,j)=1\big | |\mathbf Y^{(e)}(:,j)|_1=l_1m] }\\
& \buildrel (c) \over= \frac{Pr[|\mathbf Y^{(e)}(:,j)|_1=l_1m\big | \mathbf X^{(e)}(1,j)=0] }{Pr[|\mathbf Y^{(e)}(:,j)|_1=l_1m\big | \mathbf X^{(e)}(1,j)=1]},
\end{align}
where (c) follows due to the Bayes' expansion.
We observe that conditioned on $\mathbf X^{(e)}(1,j)=0$, $|\mathbf Y^{(e)}(:,j)|_1\sim l_1 \psi_j$, where $\psi_j\sim B(r-1, 1/2)$; and we have already seen that conditioned on $\mathbf X^{(e)}(1,j)=1$, $|\mathbf Y^{(e)}(:,j)|_1 =l_1(d +\psi_j)$. Thus, for $m\in A$,
\begin{align}
  \nonumber \texttt{ratio}(m) &= \frac{{r-1 \choose m}}{{r-1 \choose m-d}} \\
\nonumber  & = \frac{(m-d)! (r-1-m+d)!}{m!(r-1-m)!}\\
\nonumber   &\buildrel (d)\over \doteq \sqrt{\frac{(m-d)(r-1-m+d)}{m(r-1-m)}} \frac{(m-d)^{m-d} (r-1-m+d)^{r-1-m+d}}{ m^m (r-1-m)^{r-1-m}}\\
\nonumber     &\buildrel (e)\over \doteq \frac{(m-d)^{m-d} (r-1-m+d)^{r-1-m+d}}{ m^m (r-1-m)^{r-1-m}}\\
  \label{eq:636}  &= \left(\frac{m-d}{m}\right)^{m}  \left(\frac{r-1-m+d}{r-1-m}\right)^{r-1-m} \left(\frac{r-1-m+d}{m-d}\right)^d\\
\label{eq:637}& \buildrel (f) \over \doteq 1,
\end{align}
where (d) is due to the Stirling's approximation $n!=  \sqrt{2 \pi n} (n/e)^n (1 + \frac{1}{12n} +O(\frac{1}{n^2}))$ (see  \cite[p.434]{motwani95}), (e) is true since for $m\in A$, ${\frac{(m-d)(r-1-m+d)}{m(r-1-m)}} \rightarrow 1$, and (f) follows since for $m\in A$ each of the terms in \eqref{eq:636} approaches 1, since $d$ is a constant. Thus combining \eqref{eq:633}, (\ref{eq:634}), (\ref{eq:635}) and (\ref{eq:637}) implies that 
$$\frac{P_e^{maj}(\mathbf Y^{(e)})}{1-P_e^{maj}(\mathbf Y^{(e)})} \ge  1 -o(1), $$
implying $$P_e^{maj}(\mathbf Y^{(e)}) \ge 1/2 -o(1).$$
But we have already observed at the beginning of this proof that $p_0\le 1/2$. Thus $$P_e^{maj}(\mathbf Y^{(e)}) = 1/2 -o(1).$$

\noindent
{\bf Conditioned on $(C_2)$:} A very similar set of steps prove the lemma when the event $(C_2)$ occurs. The main difference is that in this case $d=O(\log n)$, unlike being a constant for $\mathbf Y^{(e)}$. But \eqref{eq:637} is still valid and hence we have $P_e^{maj}(\mathbf Y^{(e)}) = 1/2-o(1).$

\subsection{Moderate deviation for binomial distribution}
\label{app:moderate}
To prove Lemma \ref{lemma:101} and Lemma \ref{lemma:102}, we need the following theorem.
 Suppose $Q(t)$ denotes the upper tail of a standard normal distribution, i.e., $Q(t):=\frac{1}{\sqrt{2\pi}} \int_t^\infty e^{-t^2/2} dt$.
\begin{theorem}[{Moderate deviations for binomial}]
\label{thm:large}

Suppose  
$X_n\sim B(n,p_n)$. If $t_n\rightarrow \infty$ in such a way that $t_n^6=o\left(Var(X_n)\right)=o(np_n(1-p_n))$, then
$$Pr\big[X_n > np_n + t_n \sqrt{np_n(1-p_n)}\big] \doteq Q(t_n).$$
\end{theorem}
The above theorem is an adaptation of a theorem about moderate deviations of binomials when $p_n$ is a constant \cite[p. 193]{Feller1}.  The proof is very similar to the one presented in \cite{Feller1} for the constant probability case, and is omitted here.

\subsection{Hyper-geometric tails}
\label{app:hyper}
\begin{definition}[{Hyper-geometric distribution}]
\label{def:hyper}
  A random variable $X$ has hyper-geometric distribution with parameters ($N,m,n$) if 
$$h(N,m,n,t):=Pr[X=t]=\frac{{m\choose t}{N-m \choose n-t}}{{N \choose n}}, \text{ for $k=0,1,2,...,n$.}$$
It describe the number of success in a sequence of $n$ draws from a finite population, without replacement.
\end{definition}

We have the following bound for the tail of a hyper-geometric distribution, due to Chvatal \cite{Chvatal1}.
\begin{lemma}[{Hyper-geometric tail, Chvatal}\cite{Chvatal1}]
  \label{lemma:chvatal}
Suppose a random variable $X$ has hyper-geometric distribution with parameters $(N, m, n)$. Define $p:=m/N$. Then for $t\ge 0$ we have the following bound on the upper tail of $X$.
\[Pr[X \ge (p+t)n] \le \left(\left(\frac{p}{p+t}\right)\left(\frac{1-p}{1-p-t}\right)\right)^n.\]
\end{lemma}

For a hyper-geometric random variable $X$, we have $\mathbb E[X]=\frac{nm}{N}$. By observing the symmetry  $h(N,m,n,t)=h(N-m,m,n,n-t)$, we obtain the following symmetric bound for the lower tail of $X$.
\begin{corollary}[{The lower tail}]
  \label{cor:lower}
  Suppose a random variable $X$ has hyper-geometric distribution with parameters $(N, m, n)$. Define $p:=m/N$. Then for $t\ge 0$,
\[Pr[X \le (p-t)n] \le \left(\left(\frac{p}{p-t}\right)\left(\frac{1-p}{1-p+t}\right)\right)^n.\]
\end{corollary}
The following is a consequence of Lemma \ref{lemma:chvatal} and Corollary \ref{cor:lower}.
\begin{corollary}[{Simple tail bound}]
  \label{cor:tail}
Suppose a random variable $X$ has hyper-geometric distribution with parameters $(N, m, n)$. Define $p:=m/N$. Then
\[Pr\big[ \mathbb E[X](1+\delta) \le X \le \mathbb E[X](1-\delta)\big] \ge 1- 2e^{-\mathbb E[X]\delta^2/3}\]
\end{corollary}

We also need the following version of Lemma \ref{lemma:low_mean} for hyper-geometric random variables. The proof is exactly same as of Lemma \ref{lemma:low_mean}, and we refer to \cite[p.23]{Dubhashi1} for the same.
\begin{corollary}[{Tail of a hyper-geometric r.v.}]
\label{cor:low_mean_hyper}

  Suppose $X$ is a hyper-geometric random variable with parameters $(N,m,n)$, so that $\mathbb E[X]=nm/N$. For $t > 2e\mathbb E[X]$, we have
\[Pr[X > t] \le 2^{-t}.\]
\end{corollary}






\bibliographystyle{IEEEtran}
\bibliography{../../../../../../myBib}

\end{document}